\documentclass{melba}

\usepackage{mwe} 

\usepackage{amsmath,amsfonts}
\usepackage{algorithm}
\usepackage{algpseudocode}

\melbaid{YYYY:NNN}  
\doi{10.59275/j.melba.2024-AAAA}
\melbaauthors{Du, Liu, Dharmakumar and Tsaftaris}  
\email{yuning.du@ed.ac.uk}
\volume{3}
\firstpageno{1337}  
\melbayear{2026}  
\datesubmitted{yyyy-m1-d1}  
\datepublished{yyyy-m2-d2}  

\melbaspecialissue{Medical Imaging with Deep Learning (MIDL) 2020}
\melbaspecialissueeditors{Marleen de Bruijne, Tal Arbel, Ismail Ben Ayed, Hervé Lombaert}

\ShortHeadings{Active Sampling for MRI-based Sequential Decision Making}{Du, Liu, Dharmakumar and Tsaftaris}

\title{Active Sampling for MRI-based Sequential Decision Making}

\author{
	\name Yuning Du\aff{1},
	\name Jingshuai Liu\aff{1},
    \name Rohan Dharmakumar\aff{2},
    \name Sotirios A. Tsaftaris\aff{1},
}
\affiliations{
	\num 1 \addr School of Engineering, The University of Edinburgh, 5 Little France
Rd, Edinburgh, EH16 4UX, Scotland, United Kingdom \\
	\num 2 \addr School of Medicine, Indiana University, 340 West 10th Street, Indianapolis, IN
46202-3082, Indiana, United States \\
}

\abstract{
Despite the superior diagnostic capability of Magnetic Resonance Imaging (MRI), its use as a Point-of-Care (PoC) device remains limited by high cost and complexity. To enable such a future, one key approach will be to improve sampling strategies. Previous work has shown that it is possible to make diagnostic decisions directly from $k$-space with fewer samples. Such work shows that single diagnostic decisions can be made, but if we aspire to see MRI as a true PoC, multiple and sequential decisions are necessary while minimizing the number of samples acquired. 
We present a novel multi-objective reinforcement learning framework enabling comprehensive, sequential, diagnostic evaluation from undersampled $k$-space data. Our approach during inference actively adapts to sequential decisions to optimally sample. To achieve this, we introduce a training methodology that identifies the samples that contribute the best to each diagnostic objective using a step-wise weighting reward function.
We evaluate our approach in two sequential knee pathology assessment tasks: ACL sprain detection and cartilage thickness loss assessment. Our framework achieves diagnostic performance competitive with various policy-based benchmarks across disease detection, severity quantification, and overall sequential diagnosis, while substantially reducing the number of required 
$k$-space samples, thereby advancing MRI toward a more comprehensive and affordable point-of-care device.
	Our code is available at~\url{https://github.com/vios-s/MRI_Sequential_Active_Sampling}.}


\begin{document}

\twocolumn[\maketitle]

\section{Introduction}

\begin{figure}[t]
    \centering
    \includegraphics[width=0.5\textwidth]{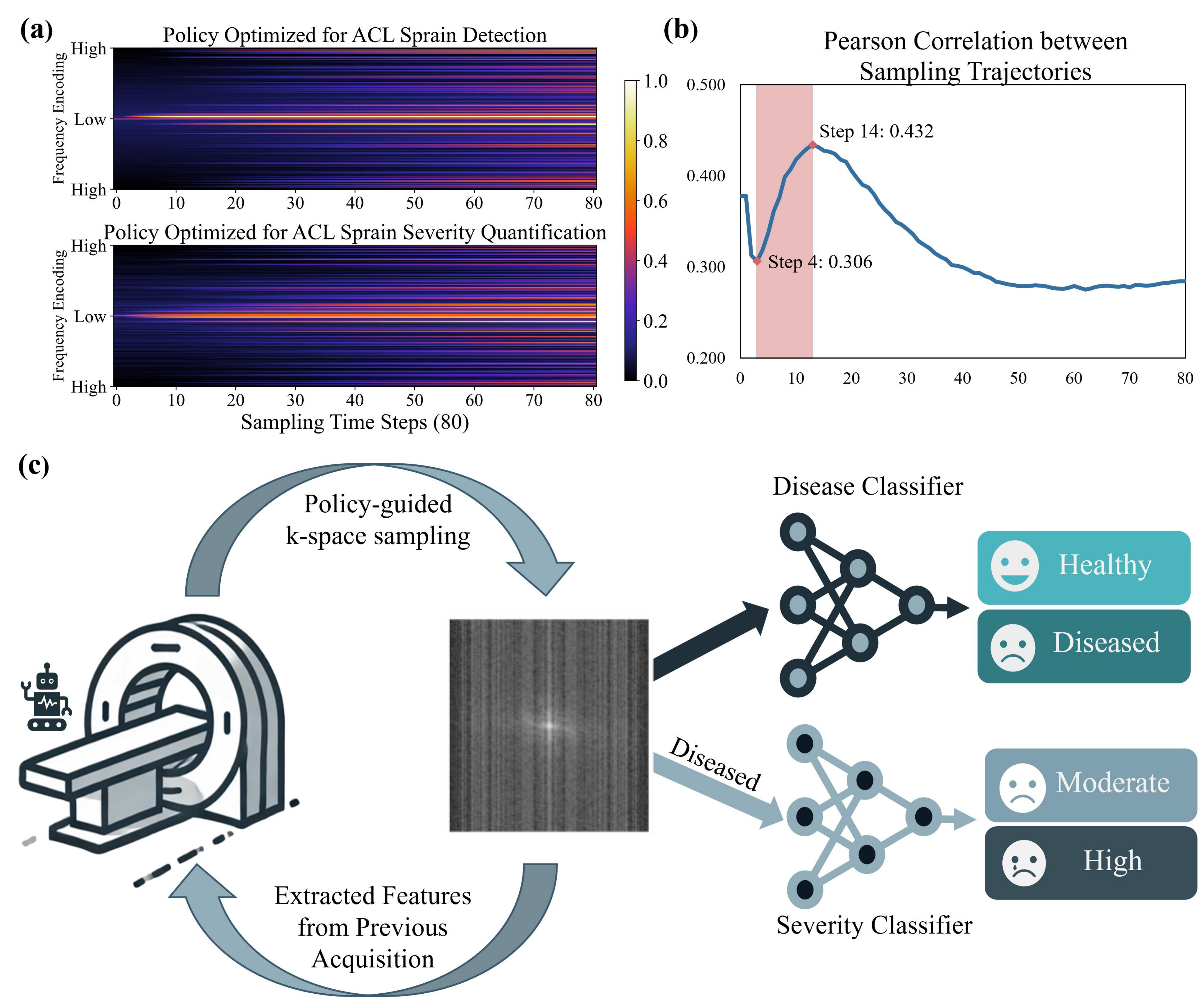}
    \caption{Illustration of our proposed sequential diagnostic framework for active MR acquisition and its motivating evidence. In preliminary experiments, we found similarities between the sampling trajectories of disease detection and severity quantification. We have trained policies~\citep{du2024mri} optimized independently for each task and obtained sampling trajectories shown in (a), and then measure their similarity in (b). We observe an increasing Pearson correlation between sampling steps $4$ and $14$, indicating that in this range $k$-space samples contribute simultaneously to both tasks. These observations motivate our approach (c) to learn a sampling policy that jointly optimizes two tasks, taking into account their sequential nature. During inference, our approach identifies which samples to take using $k$-space features, considering that one task (severity) will be following the confirmation of the disease. 
    }
    \label{fig: proof of concept}
\end{figure}

\enluminure{D}{espite} the proliferation of MRI, its utility as a Point-Of-Care (PoC) diagnostic tool remains limited due to prolonged acquisition times, complex instrumentation, and high cost~\citep{geethanath2019accessible}. A promising solution has emerged through the development of low-field MRI systems~\citep{anzai2022point}. These systems are being evaluated for bedside and point-of-care use~\citep{anzai2022point}, where immediate decisions are necessary to provide the best treatment options. Such PoC systems are not intended to replace MRI as a general clinical diagnostic tool, but to serve as bespoke, affordable and portable devices. However, a reduced magnetic field strength~\citep{hori2021low} does degrade imaging quality. Therefore, improvements in how and where samples are taken are considered key for unlocking the future of MRI as PoC.

Recent advances in machine learning methods have allowed $k$-space samples to be reduced while retaining diagnostic utility~\citep{lin2021artificial}. A line of work has focused on optimising sampling trajectories for image reconstruction~\citep{bahadir2020deep,zhang2019reducing}. More recently, approaches have emerged that aim to reduce sampling substantially by removing redundant information. These methods bypass image reconstruction entirely and make direct inferences from undersampled $k$-space~\citep{singhal2023feasibility,yen2024adaptive}. 

Decisions even in a PoC setting will be complex and reflect a multilayered decision-making process. Hence, high-level decisions (e.g., disease presence) should be made first, followed by additional decisions (e.g., disease severity) to determine subsequent actions (e.g., treatment options). A fundamental research question emerges: \textit{Are there correlations in $k$-space patterns between such sequential decisions and can these be exploited for further sampling reductions?}  
If such correlations exist, sampling strategies should be optimised to exploit them. Otherwise, independently optimising sampling strategies for each diagnostic task leads to sampling inefficiency. Preliminary experiments (Figure~\ref{fig: proof of concept}) reveal consistent correlation patterns between sampling trajectories.

Leveraging this observation, this chapter develops an AI agent to satisfy both objectives within the same policy, maintaining computational efficiency while structuring binary diagnostics as a structured sequential decision process that assesses both disease presence and severity. To the best of our knowledge, such active sampling strategies for sequential diagnostic tasks have not been previously explored. This chapter reformulates the diagnostic process as a multi-objective reinforcement learning (MORL) sampling optimisation problem, sequentially inferring the presence of the disease and its severity in one scanning session. 

As illustrated in Figure~\ref{fig: proof of concept},  the core components of the proposed framework include: an RL-based policy network for $k$-space sampling guidance, and dual diagnostic streams for disease detection and severity quantification. 
The policy network guides data acquisition by leveraging features extracted from accumulated $k$-space measurements, continuously optimising the sampling trajectory. 
The diagnostic streams process the undersampled $k$-space hierarchically: the disease classifier performs binary classification to identify diseased subjects, which is followed by severity quantification for positive cases. 

Overall, the \textbf{contributions} of this chapter are summarised as follows:

\begin{enumerate}
    \item[$\bullet$] This chapter introduces sequential, dynamic diagnosis from $k$-space data with active sampling, providing sequential decisions in a single scanning session.
    \item[$\bullet$] A policy gradient based network with greedy action selection enables efficient $k$-space exploration and patient-level sampling adaptation. 
    \item[$\bullet$] A step-wise weighting reward function is proposed to balance disease detection and severity quantification during policy training, maximising diagnostic information gain.
    \item[$\bullet$] Extensive experiments on ACL sprain detection and cartilage thickness loss assessment demonstrate that the proposed method achieves the highest AUC among undersampled baselines and benchmarks in most settings, and substantially reduces $k$-space data acquisition time, advancing PoC MRI applications.
\end{enumerate}


\section{Related Works}
\label{previouswork}

Three bodies of work frame the proposed method: ML-based $k$-space sampling optimisation, direct diagnostic inference from $k$-space, and RL-based multi-objective optimisation in medical imaging. None of them addresses sequential decisions within one acquisition, which is the gap the proposed framework fills by diagnosing disease presence and severity directly from undersampled $k$-space.

\subsection{ML-based \texorpdfstring{$k$}{k}-space Sampling Optimisation}
\label{RW:samplingoptim}

Accelerating MRI acquisition remains a significant research focus through improved $k$-space undersampling strategies. While the Cartesian grid trajectory predominates in clinical practice due to its robustness~\citep{zeng2021review}, conventional undersampling approaches, including equispaced, random, and variable density sampling, prioritise frequency domain characteristics over task-specific requirements, potentially compromising clinical accuracy in accelerated acquisitions~\citep{yiasemis2024retrospective}. 

Recent works have advanced learning-based sampling methods that optimise sampling at the population level. These approaches formulate sampling mask optimisation either as a probability density function for Cartesian grid selection~\citep{bahadir2020deep,wang2022joint,martinini2022deep,aggarwal2020j} or as learnable non-Cartesian coordinates~\citep{zhang2020extending,chaithya2022hybrid} in non-uniform Fast Fourier Transform (NUFFT)~\citep{wang2022b,wang2023efficient,alkan2020learning}, enabling end-to-end optimisation of sampling trajectories and reconstruction performance.  Advancements have also been made in co-design frameworks that jointly optimise sampling, reconstruction, and prediction strategies for enhanced performance on downstream tasks~\citep{wu2024learning}. While subsequent innovations incorporate generative models~\citep{ravula2023optimizing} and model pruning~\citep{xuan2020learning}, these methods produce fixed sampling patterns that do not adapt to patient-specific characteristics. 

Instead, active sampling strategies adapt to individual patients through sequential $k$-space measurements. Reinforcement learning (RL) is well suited to this sampling optimisation, framing $k$-space sampling as a Markov Decision Process where policies learn to select measurements based on accumulated information. Notable approaches include policy gradient methods with greedy actions~\citep{bakker2020experimental} and Q-learning~\citep{pineda2020active} for reconstruction optimisation. While initial work focused on reconstruction quality, recent studies demonstrate the potential for direct diagnostic inference from undersampled data~\citep{schlemper2018cardiac,li2024classification}, indicating a move from reconstruction-based analysis towards direct diagnostic assessment from $k$-space.

\subsection{Direct Diagnostic Inference from \texorpdfstring{$k$}{k}-space}
\label{RW:DirectInference}
A new line of work has shown the potential of diagnostic inference on undersampled spatial representations, avoiding a reconstruction-first pipeline while operating on the inverse-Fourier transform of the undersampled $k$-space rather than on raw frequency-domain measurements directly. This approach reduces the need for reconstruction, saving samples by avoiding redundant and uninformative pixels (voxels). Studies show successful disease classification~\citep{du2024mri,singhal2023feasibility,yen2024adaptive,li2024classification}, segmentation~\citep{schlemper2018cardiac,li2024classification}, and regression~\citep{li2024classification} on undersampled spatial representations, without a separate reconstruction stage. Direct inference with population-level undersampling optimisation achieves comparable performance to fully-sampled image analysis, while integration with RL-based active sampling enables patient-level mask optimisation and improved classification performance~\citep{du2024mri,yen2024adaptive}. However, existing approaches primarily optimise sampling for a single objective during a scanning session, leaving multi-objective sampling optimisation unexplored. As such, any additional redundancy between the sampling patterns for different tasks (and hence different objectives) is unutilised. 

\subsection{RL-based Multi-objective Optimisation in Medical Imaging}
\label{RW:MORL}
Multi-objective optimisation in clinical evaluation systems presents a fundamental challenge in medical imaging. While multi-task learning frameworks have been explored extensively~\citep{zhao2023multi,huang2021deep}, the joint optimisation of sampling strategies for multiple diagnostic targets remains underexplored. 
Recent advances in RL show promising methodological innovations across medical applications~\citep{yu2021reinforcement}. In chest X-ray generation~\citep{han2024advancing}, policy-based RL simultaneously optimises posture alignment, diagnostic conditions, and multimodal consistency through cumulative reward maximisation. Similarly, anatomical landmark detection~\citep{vlontzos2019multiple} leverages multi-agent frameworks with specialised joint detection while preserving anatomical constraints. Further innovations include consequence-oriented reward functions for simultaneous vertebral detection and segmentation~\citep{zhang2021sequential}, incorporating exponentially weighted attention metrics for balanced task performance. 

While various frameworks demonstrate efficacy through weighted rewards
~\citep{zhang2021sequential} and collaborative strategies~\citep{vlontzos2019multiple,liao2020iteratively}, their application to $k$-space sampling optimisation remains unexplored, presenting opportunities for frameworks that simultaneously optimise multiple diagnostic goals.

\section{Methodology}
\label{Method}
\subsection{Preliminaries}
Instead of directly imaging human anatomy, MRI captures the electromagnetic activity in the body after exposure to magnetic fields and radio-frequency pulses, measured in $k$-space (i.e., the frequency domain). For a single coil measurement, the $k$-space data can be represented as a 2-dimensional complex-valued matrix $\mathbf{x}\in \mathbb{C}^{r \times c}$, where $r$ is the number of rows and $c$ is the number of columns, respectively. The spatial representation $\mathbf{SR}\in \mathbb{C}^{r \times c}$ is obtained by applying the inverse Fourier Transform to $\mathbf{x}$, denoted as $\mathbf{SR} = \mathcal{F}^{-1}(\mathbf{x})$, preserving both the magnitude and phase information. The spatial image $\mathbf{I}\in \mathbb{R}^{r \times c}$  retains only the magnitude of $\mathbf{SR}$, computed as $\mathbf{I} = |\mathbf{SR}|$, to provide an interpretable representation in image space. The undersampled $k$-space, represented as $\mathbf{x}_u$, is modeled by $\mathbf{x}_u = U_L\circ \mathbf{x}$ where $U_L$ can be viewed as a binary mask $U \in \{0, 1\}^{r \times c}$ with $L$ measurements sampled in $k$-space. The undersampled representation is denoted as $\mathbf{SR}_u = \mathcal{F}^{-1}(\mathbf{x}_u)$.

We exclusively consider a Cartesian mask for MRI undersampling. While non-Cartesian trajectories (e.g., radial, spiral) can offer more efficient $k$-space coverage, the Cartesian grid was chosen for its widespread clinical prevalence and methodological simplicity~\citep{reina2025mrilow}. This choice avoids the complex gridding steps required by other types of trajectories which could introduce confounding artifacts, and thereby ensures a clear and direct evaluation of our adaptive sampling policy.

For clinical diagnosis, radiologists and other clinical experts interpret medical imaging data to first identify the presence of disease ($g_d$) and then its severity ($g_s$), providing a comprehensive diagnosis. These labels are derived by analyzing the collected $k$-space data after appropriate processing.

\begin{figure}[t]
    \centering
    \includegraphics[width=1.0\linewidth,keepaspectratio]{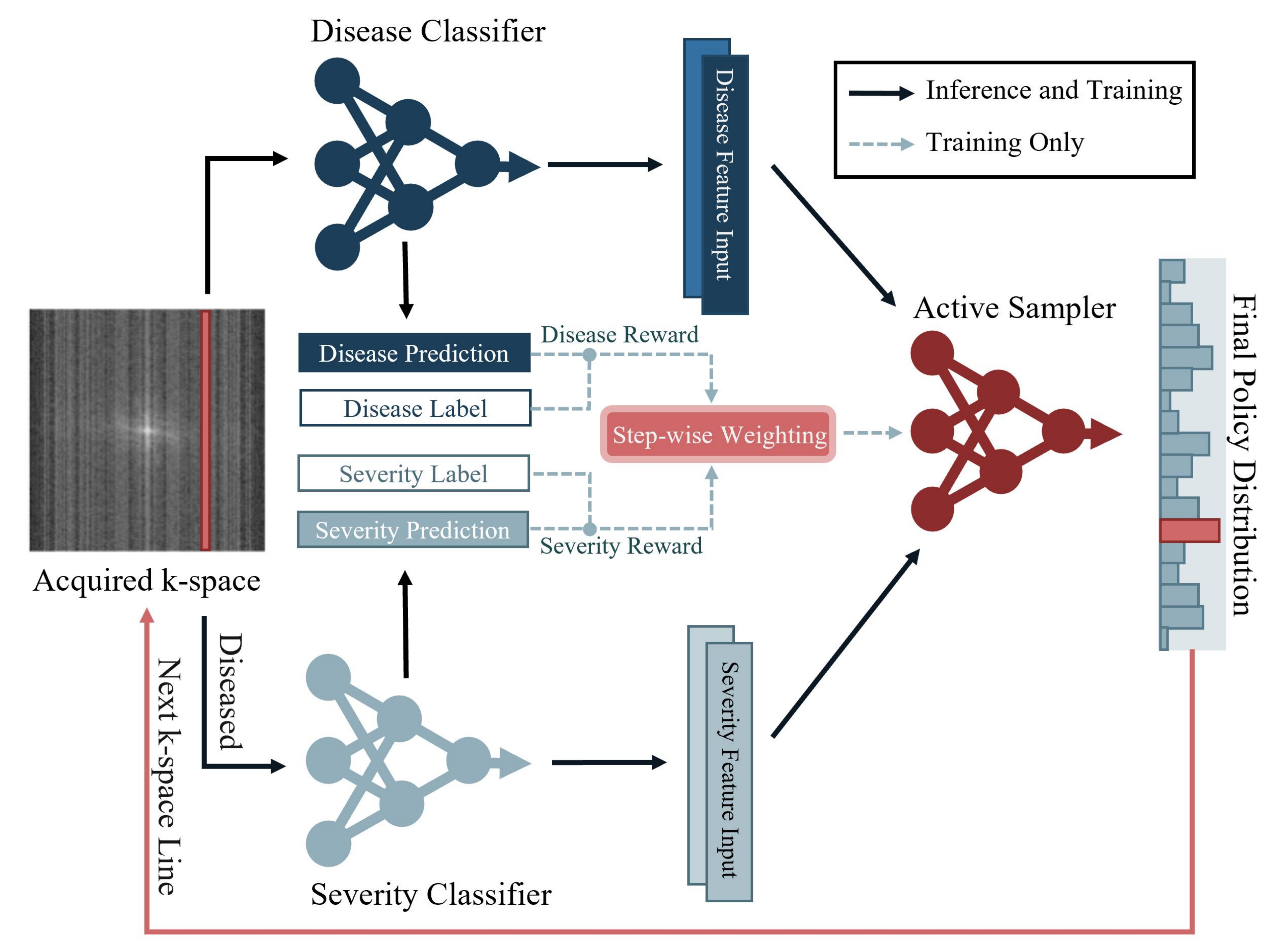}
    \caption{Illustration of the proposed sequential diagnostic active sampling. Initially, a randomly sampled $k$-space subset is fed into two pre-trained disease and severity classifiers. The active sampler takes as input the features extracted by the classifiers and decides the next $k$-space sampling location. After exhausting the sampling budget or meeting a user-defined criterion, the disease classifier confirms the presence of disease and subsequently the severity classifier is queried to estimate the severity level. During training, the active sampler is optimized with a novel step-wise weighting mechanism that combines the disease and severity rewards and enables a smooth transition.} 
    \label{fig framework}
\end{figure} 

\subsection{Framework Workflow}
As shown in Figure~\ref{fig framework}, our framework aims to reduce acquisition time by selectively and progressively sampling the $k$-space, considering sequential diagnostic tasks.

The process begins with a small subset of randomly sampled $k$-space with $L$ measurements, denoted as $\mathbf{x}_{u_0} = U_L\circ \mathbf{x}$, which is passed to task-specific classifiers. The pre-trained disease and severity classification networks $f_{d}$ and $f_{s}$, generate the initial disease and severity prediction $y_{d_0}$ and $y_{s_0}$. 
The high-level feature maps $m_0$, from the last two layers of both classifiers, are extracted and stacked. The feature maps are then fed into the \textit{Active Sampler} $S$, where the policy network, parameterized by $\phi$, generates a sampling policy $\pi_\phi(m_0)$ to guide the selective sampling of diagnostically significant lines in the $k$-space. 
These sampled lines are subsequently added to $U_L$. The updated $k$-space subset at step $t \in [0, T]$ is denoted as $\mathbf{x}_{u_t} = U_{t}\circ \mathbf{x}$ and fed back into the classification network. 
The iterative process continues until the sampling budget $T$ is exhausted or some user-defined reliability criterion (e.g., confidence score) is satisfied. The closed-loop design of the inference stage is presented in Algorithm~\ref{alg: closed loop inference}.

During training, we separately calculate the disease reward $r_{d}$ and severity reward $r_{s}$ using classifier predictions and the ground truth labels. An obvious and naive approach would be to directly use the rewards to supervise the active sampler; however, this will not learn the sequential aspect of the two objectives. Instead, we propose here to use a step-wise weighting procedure to shift the emphasis of sampling strategy from one objective to the other as sampling steps progress. In addition, to further enforce the sequential nature (i.e., disease prediction should precede severity), if the disease prediction is `No findings' for a subject, the severity reward will not be applied for supervision.

\subsection{Undersampled \textit{k}-Space Classification Networks}
To perform prediction with undersampled $k$-space data, we pre-train a disease classifier $f_{d}$ with the undersampled representation derived from  undersampled $k$-space data and ground truth label $g_{d}$. For the severity classifier $f_{s}$, we fine-tune the disease classifier with only diseased data and the associated severity labels $g_{s}$. For severity quantification, fine-tuning prevents potential overfitting caused by data scarcity and reuses the knowledge learned by the disease classifier. These pre-trained classifiers act as the reward model in the training process of the active sampler. They also function as a feature extractor, providing complex and informative feature maps $m_t$ per sampling step as part of the internal state.

\begin{algorithm}[t]
{\fontsize{9pt}{11pt}\selectfont
\caption{Closed-loop Active Sampling Inference}
\label{alg: closed loop inference}
\begin{algorithmic}[1]
\Require Pretrained disease classifier \(f_d\), pretrained severity classifier \(f_s\), policy \(\pi_\phi\), initial Cartesian mask \(U_L\) with \(L\) lines, sampling budget \(T\), underlying $k$-space \(x\)
\State \(x_u^{0} \gets U_{0} \circ x\) with \(U_{0} \gets U_L\)  \Comment{Acquire initial subset} 
\State \((y_d^{0}, \, y_s^{0}) \gets \big(f_d(x_u^{0}), \, f_s(x_u^{0})\big)\)  \Comment{Initial predictions}
\State \(m^{0} \gets \textsc{ExtractFeatures}(f_d, f_s)\)   \Comment{High-level feature maps from both streams}
\For{\(t = 0, 1, \dots, T-1\)}
    \State \(p \gets \pi_\phi(m^{t})\)  \Comment{Policy over High-level feature maps}
    \State \(a_t \gets \arg\max\limits_{\ell \in \mathcal{L} \setminus U_t} \, p[\ell]\) \Comment{Select next line (inference: 1 line/step)}
    \State \(U_{t+1} \gets U_t \cup \{a_t\}\)  \Comment{Update mask}
    \State \(x_u^{t+1} \gets U_{t+1} \circ x\)  \Comment{Acquire measurement at line \(a_t\)}
    \State \(y_d^{t+1} \gets f_d(x_u^{t+1})\); \quad \(y_s^{t+1} \gets f_s(x_u^{t+1})\)  \Comment{Update predictions}
    \State \(m^{t+1} \gets \textsc{ExtractFeatures}(f_d, f_s)\)   \Comment{Update internal state}
\EndFor
\State \(\hat{y}_d \gets y_d^{T}\)
\If{\(\textsc{IsPositive}(\hat{y}_d)\)}
    \State \(\hat{y}_s \gets y_s^{T}\)  \Comment{Severity only reported for disease-positive cases}
\Else
    \State \(\hat{y}_s \gets \textsc{NotApplicable}\)
\EndIf
\State \Return \(\hat{y}_d, \hat{y}_s\)
\end{algorithmic}
}
\end{algorithm}

\subsection{Multi-objective \textit{k}-Space Active Sampler}

\subsubsection{Problem Formulation}
Building on prior works in RL-based $k$-space active sampling strategies~\citep{du2024mri,bakker2020experimental,pineda2020active}, the sequential selection of $k$-space is formalized as a Partially Observable Markov Decision Process (POMDP)~\citep{sondik1971optimal}.

The latent state $z_t$ at acquisition step $t$ is defined as $z_t = (\mathbf{x}, g_d, g_s, h_t)$, where $\mathbf{x}$ represents the underlying MRI $k$-space, $g_{d}$ and $g_{s}$ denote the disease condition and severity levels, and $h_t$ encapsulates the history of actions and observations up to time $t$ as $h_t = (a_0,o_0,...,a_{t-1},o_{t-1})$. 
Each action $a_t$ corresponds to the selection of a specific $k$-space line for measurement, while $o_t$ represents the corresponding measurement obtained from that location. The inclusion of $\mathbf{x}$ in the latent state represents the ground truth data from which $k$-space measurements are derived, fundamentally linking the observation model to the true anatomical structure.
The agent maintains an internal state composed of high-level features $m_t$ that combine disease and severity features from the corresponding classifiers, and predictions $y_t$ including disease prediction $y_{d_t}$ and severity prediction $y_{s_t}$, which summarize the history $h_t$. The agent outputs a policy $\pi(a_t|m_t)$ over possible $k$-space line selections with only high-level feature representation as input. The reward function $r(z_t, a_t)$ reflects improvements in both disease classification and severity quantification accuracy, weighting their respective contributions through a step-wise mechanism, as discussed below.

\subsubsection{Policy Gradient-based Sampler with Greedy Action}

In our framework, the POMDP problem is solved via Policy Gradient with greedy action. Specifically, we maximize the expected return $J(\phi)$ of a policy $\pi_\phi$ parameterized by $\phi$. At each step, the overall classification improvement is calculated as $r(y_{t+1}, y_t) = \eta(y_t, g) - \eta(y_{t+1}, g)$, where $r$ represents the reward computed with criterion $\eta$ (cross entropy) between predictions $y$ and ground truth labels $g = (g_d, g_s)$.

During training, we accelerate the learning process by sampling $q$ lines with the highest policy probabilities in parallel. The greedy selection method facilitates efficient action space exploration. Technically, we average their accumulated action rewards as feedback for model optimization~\citep{kool2019buy} and randomly select one of the $q$ samples as the executed action of the policy. As no greedy action is applied during inference, the trained policy samples one line at each step.   
We sample $q$ lines at each step to calculate the following estimator:
\begin{equation}\label{eq1}
\resizebox{0.5\textwidth}{!}{
  $\displaystyle
  \begin{aligned}
  \nabla_\phi J(\phi) \approx & \frac{1}{q-1} \mathbb{E}_{\mathbf{x}}[ \sum_{i=1}^q \sum_{t=L}^{T-1} \left[\nabla_\phi \log \pi_\phi\left(m_t\right)\right] (r_{i, t} - \frac{1}{q} \sum_{j=1}^q r_{j, t})], 
  \end{aligned}$
}
\end{equation}
where $r_{i, t}$ is the reward of sample $i$ at step $t$. This equation estimates the gradient $\nabla_\phi J(\phi)$ to update the policy $\pi_\phi$.

\subsubsection{Step-wise Reward Weighting with Lexicographic Ordering}
\label{sec: weighted sum reward}
\begin{figure}[ht]
    \centering
    \includegraphics[width=1.0\linewidth]{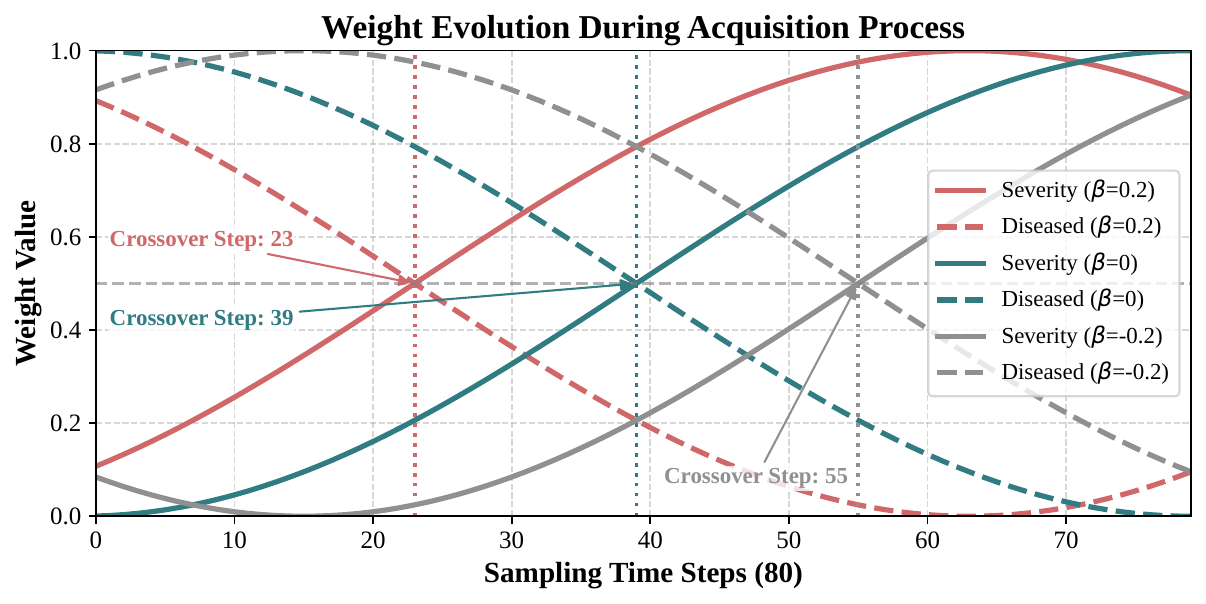}
    \caption{Step-wise weighting evolution for multi-objective diagnosis following lexicographic ordering. The step-wise weight  dynamically determines the preference of the reward feedback in disease detection and severity quantification objectives, which ensures a smooth transition. Different $\beta$ values enable different transition patterns, which can be adjusted based on optimized objectives .}
    \label{fig: adaptive weight}
\end{figure}

As mentioned in Sec.~\ref{RW:MORL}, a crucial challenge in multi-objective $k$-space sampling is to jointly optimize multiple objectives without compromising performance. One naive solution is to respectively compute the rewards for the two tasks (here disease detection and severity quantification) and use the total reward as feedback for policy update. However, this ignores the natural order of these two tasks: one first detects disease presence and only if disease is present, the degree of severity is quantified, which aligns with the standard interpretation workflow of radiologists~\citep{royal2018standards}. Following such diagnostic order, we expect the sampling framework to primarily promote disease detection performance in the early stage and focus on severity quantification in the later stage. Consequently, to accomplish these tasks with MORL, we formulate the process as a lexicographic ordering problem~\citep{skalse2022lexicographic}, where the order of the two objectives is fixed and the weight is dynamically adjusted by \textit{cosine interpolation}~\citep{cosineinterpolation} during the sampling process to ensure a smooth transition. Thus, we define a step-wise weighting reward with cosine interpolation for each sample as follows:
\begin{subequations}
    \begin{align}
        &r_t = w_{disease}(t)*r_{d}(t) + w_{severity}(t)*r_{s}(t) \label{eq:reward}\\
        &w_{severity}(t) = \frac{1}{2}(1 - \cos(\pi \cdot \frac{t + 1}{T} + \pi\beta)) \label{eq:w_severity}\\
        &w_{disease}(t) = 1 - w_{severity}(t), \label{eq:w_disease}
    \end{align}
    \label{eq:reward_severity_disease}%
\end{subequations}
where $r_d$ and $r_s$ denote disease detection and severity quantification rewards, $t$ is the current step, $T$ is the total acquisition steps, and $\beta$ denotes a weight parameter which adjusts the equal weighting point. Here we set $\beta$ to be $0.2$ to enable an early transition.
The transition of the two objectives is also illustrated in Figure~\ref{fig: adaptive weight}. It is shown that by step-wise weighting, the reward function emphasizes the feedback of the disease detection task in the early stage and smoothly transitions to severity quantification. Different $\beta$ values enable different transition patterns, which can be adjusted based on the complexity and importance of the two optimized objectives.   

\begin{table*}
    \centering
    \caption{Objectives, Diagnostic Support, and Inference Data of Implemented Methods. The objective of Policy Recon is image reconstruction}
    \label{tab: diagnostic support}
    \resizebox{1.0\textwidth}{!}{\begin{tabular}{|c| c c c| c c| c c|} 
    \hline
     & \multicolumn{3}{|c|}{Diagnostic Objectives} & \multicolumn{2}{c|}{Diagnostic Strengths} & \multicolumn{2}{c|}{Data Used for Inference } \\
    \hline
    Method & Disease Detection & Severity Quantification & Overall Diagnosis & Sampling Optimization & Sequential Diagnosis & Undersampled $k$-space & Full $k$-space \\
    \hline
    \textit{Fully Sampled}  & $\circ$ & $\circ$ & $\circ$ &  &  &  & $\checkmark$ \\
    \textit{Undersampled} & $\circ$ & $\circ$ &  &  &  & $\checkmark$ & \\
    \hline
    \textit{Policy Recon~\citep{bakker2020experimental}}  & - & - & - & $\checkmark$ &  & $\checkmark$ & \\
    \textit{Policy Classifier~\citep{du2024mri}} & $\circ$ & $\circ$ &  & $\checkmark$ &  & $\checkmark$ & \\
    \hline
    \textit{Simulated Policy} &  &  & $\circ$ & $\checkmark$ &  & $\checkmark$ & \\
    \textit{Varying Parameter Policy} & $\bullet$ & $\bullet$ &  & $\checkmark$ &  & $\checkmark$ & \\
    \hline
    \textit{Random Policy}  & - & - & - &  &  & $\checkmark$ & \\
    \textbf{\textit{Weighted Policy (Ours)}} & $\bullet$ & $\bullet$ & $\bullet$ & $\checkmark$ & $\checkmark$ & $\checkmark$ & \\
    \hline
    \multicolumn{8}{|l|}{\footnotesize{\; $\circ$ : Objective optimized independently (one objective per strategy)}
    \footnotesize{$\bullet$ : Multiple objectives optimized simultaneously within a single strategy}} \\
    \hline
    \end{tabular}
    }
\end{table*}

\section{Experimental Setup}
\label{sec:setup}
\subsection{Dataset and Pre-processing}
\label{sec:Data}
\noindent \textbf{Dataset}:
Experiments are conducted with the single-coil $k$-space data and slice-level labels from the publicly available fastMRI dataset~\citep{zbontar2018fastmri} and fastMRI+ dataset~\citep{zhao2022fastmri+}. We use $1,172$ annotated volumes from the fastMRI Knee dataset to validate the performance of our framework in two types of pathology diagnosis and severity quantification: \textit{ACL Sprain (High Grade/Low-Moderate Grade)} and \textit{Cartilage Thickness Loss (Full/Partial)}. We follow the data split provided by the dataset, resulting in the distribution of Table ~\ref{tab:label_distribution}.

\begin{table}
\centering
\caption{Train-Val-Test Split for ACL Sprain and Cartilage Thickness Loss datasets}
\label{tab:label_distribution}
\resizebox{0.5\textwidth}{!}{
\begin{tabular}{|c| c c c| c c c|} 
\hline
Pathology  & \multicolumn{3}{c|}{ACL Sprain} & \multicolumn{3}{c|}{Cartilage Thickness Loss} \\
\hline
Severity Degree & No Finding & Low-Mod Grade & High Grade & No Finding & Partial & Full \\
\hline
Train & $30915$&$617$& $503$& $29768$& $1850$& $417$\\
Validation & $5180$&  $56$& $58$&  $4907$& $325$& $62$\\
Test &  $1734$& $42$& $65$& $1732$& $79$& $30$\\
\hline
Total & $37829$ & $715$ & $626$ & $36407$&  $2254$&  $509$\\
\hline
\end{tabular}
}
\end{table}

\noindent \textbf{Data Pre-processing}:
Since the $k$-space data have varying sizes, we first apply inverse Fourier transform to obtain the fully sampled $k$-space data to get the ground truth image and crop it to size $(320 \times 320)$ for computational efficiency. After applying Fourier transform to the ground truth image, we obtain the ground truth  $k$-space data.
There is a severe class imbalance between the two diagnostic tasks. To address it during training, we use weighted cross-entropy loss based on the label distribution to avoid overfitting to the majority class and poor generalization on the minority class. When training the policy network, we undersample the data for computational savings, as suggested in~\citep{bakker2020experimental,du2024mri}.  

\subsection{Baselines and Benchmarks}
\label{sec: baselines}

We design benchmark methods and baselines based on different diagnostic strengths, data access, and objectives as shown in Table~\ref{tab: diagnostic support}, to fairly evaluate our approach. 

\subsubsection{Fully Sampled Classifier (Oracle)}
This serves as an oracle benchmark for classifier performance with fully sampled $k$-space data as input, and hence no sampling optimization occurs. We trained the disease classifier with the ground truth spatial images $\mathbf{I}$ which are derived from fully sampled $k$-space data and supervised by the disease level label. 
The severity classifier is fine-tuned from the well-trained disease classifier with only diseased subjects and their severity label. 
The sequential diagnosis results are inferred by connecting disease and severity classifiers together. We refer to it as \textit{Fully Sampled}.

\subsubsection{Undersampled Classifier} 
The classifier serves as a baseline for single binary tasks, such as disease diagnosis and severity quantification, with spatial representation $\mathbf{SR}$ which derived from the randomly undersampled $k$-space data and no sampling optimization. It adopts the same backbone as the \textit{Oracle} and is trained with undersampled representations following Variable Density Sampling with various sample rates ($5\%$ to $30\%$) and center fraction ($0\%$ to $5\%$), supervised as before. We refer to it as \textit{Undersampled}.

\subsubsection{Single-task Oriented Methods}
The methods introduced here are designed for single-task evaluation, targeting binary diagnosis of disease detection or severity quantification.

\noindent \textbf{Policy Classifier~\citep{du2024mri}:}  
This diagnosis-oriented policy benchmark optimizes patient-level sampling in $k$-space for direct inference without reconstruction, setting the upper bound for disease detection and severity quantification. For disease detection, we use a backbone similar to \textit{Undersampled} as a pre-trained classifier. For severity quantification, the classifier is fine-tuned from the disease detection model. Sampling policy rewards are computed via weighted cross-entropy predictions, with feature maps from the pre-trained classifier as input to the policy network. Training starts with a $5\%$ initial sample rate following Variable Density Sampling, progressively sampling $80$ lines to reach a $30\%$ budget.

\noindent \textbf{Policy (via) Reconstruction~\citep{bakker2020experimental}:}  
This model optimizes patient-level sampling with image reconstruction error as rewards, providing a comparison when reconstruction is involved in diagnosis. We pre-train a U-Net~\citep{ronneberger2015u} with a $16$-channel first feature map and $4$ pooling cascades, totaling $837\mathrm{K}$ parameters. The model is trained across sampling rates ($5\%$–$30\%$) with a $5\%$ center fraction using Variable Density Sampling, supervised using $\ell_1$ loss. The active sampler, built on the reconstruction model, uses reconstructed images as input and optimizes sampling with SSIM~\citep{wang2004image} as the reward. This cascaded design totals $26.7\mathrm{M}$ parameters. Its diagnostic evaluation adopts the \textit{Oracle} classifiers, which are trained to make predictions from fully sampled spatial images, leading to the Policy-based Reconstruction Network (\textit{Policy Recon}).

\subsubsection{Multi-task oriented Methods for Comprehensive Diagnosis}

These diagnosis-oriented methods are designed for multi-task evaluation, aiming to provide a comprehensive diagnosis of the presence and severity of disease. These methods do not consider the order of objectives and enforce sequential steps.

\noindent \textbf{Simulated Policy Classifier:} This benchmark evaluates a single-objective reinforcement learning (SORL) policy network tailored for comprehensive diagnosis. It adopts a similar architecture to our framework, e.g., extracting the high-level features from the pre-trained disease and severity classifiers, while using a different rewarding mechanism. Instead of relying on step-wise weighting to adjust the diagnostic rewards, this policy is optimized via a reward function that unifies the diagnostic outcomes as the F1-score of a three-class classification task, where the final diagnosis is categorized as: (0) no findings, (1) diseased with low severity, and (2) diseased with high severity. This formulation simulates the situation where the model simultaneously infers both the presence and severity of disease, without enforcing sequential steps. We refer to this approach as \textit{Simulated Policy}.

\noindent \textbf{Varying Parameter Policy Classifier:} To benchmark diagnosis using multi-objective reinforcement learning (MORL) without specifying the optimization order of objectives, we introduce a method based on the varying parameter approach~\citep{liu2014multiobjective}. This method employs two separate policy networks—one for disease detection and another for severity quantification—designing sampling trajectories independently before averaging their outputs to guide $k$-space selection. Each policy is trained with its individual reward, but during joint training, the severity policy considers rewards only from diseased subjects, ignoring others. This approach is termed \textit{Varying Parameter Policy}.

\subsubsection{Weighted Sequential Policy Classifier (Ours)} Our proposed method, introduced in Section~\ref{Method} is a MORL-based policy network coupled with a step-wise weighting mechanism. The pre-trained disease classifier shares the same backbone as \textit{Undersampled} and only has access to undersampled $k$-space data. The severity classifier is fine-tuned from the well-trained disease classifier. Both classifiers are utilized to provide step-wise weighting rewards and high-level feature representations to the policy network as illustrated in Figure~\ref{fig framework}. During joint training, only when the input subject is diseased does the severity reward from the classifier get used. Our method is referred to as \textit{Weighted Policy}.

To establish a direct baseline for our method, we include the \textit{Random Policy}. This baseline utilizes a randomly initialized network identical in architecture to the \textit{Weighted Policy} to provide unguided, stochastic selection for the next sampling location.

\subsection{Evaluation Metrics}

To assess diagnostic performance, we evaluate our method from three perspectives: disease detection, severity quantification, and overall sequential diagnosis, using \textbf{Area Under the Receiver Operating Characteristic Curve (AUC)} and \textbf{Balanced Accuracy} as primary metrics.  
For disease detection, we assess the model’s ability to distinguish pathological cases from no findings. Severity quantification focuses on positive cases, evaluating accuracy in classifying disease progression (Low-Moderate vs. High Grade for ACL Sprain; Partial vs. Full for Cartilage Thickness Loss).  
Sequential diagnosis treats the process as trajectory prediction, requiring correct classification at both stages—detecting disease and assessing severity—for a prediction to be valid, aligning with real-world clinical needs. Additionally, sampling and computational efficiency metrics have been included for a comprehensive evaluation.

Beyond standard diagnostic performance, we incorporate four metrics that specifically address clinical risk and patient safety across the diagnostic pipeline. For disease detection, we focus on the risks associated with false negatives. The \textbf{disease detection missed rate} evaluates the proportion of diseased subjects misclassified as healthy, a critical error that can delay necessary treatment. Building on this, we specifically isolate the number of \textbf{missed high-severity subjects}. This metric highlights a subset of patients for whom a missed diagnosis introduces significantly higher clinical risk.
For severity quantification stage, we examine misclassification errors among diagnosed patients. We first assess \textbf{low-severity overestimation rate}, where incorrectly escalating the severity level may lead to wasted medical resources. Conversely, we track \textbf{high-severity underestimation rate}, an error that carries the dangerous consequence of potential undertreatment.

\subsection{Implementation Details}

All classification networks employed share the same ResNet-50~\citep{he2016deep} backbone with $25.6 \mathrm{M}$ parameters. To avoid overfitting caused by class imbalance, a dropout layer is added at the end, and weighted cross-entropy is used as the supervised learning loss~\citep{pmlr-v202-mao23b}. 
For methods involving a policy with greedy actions, we use the policy gradient network ($11 \mathrm{M}$ parameters)~\citep{du2024mri,bakker2020experimental} with parallel acquisition $q = 8$. The network consists of convolutional feature extraction layers with instance normalization followed by a three-layer fully connected head that outputs action logits for $k$-space line selection.  
For all methods, we employ the Adam optimizer~\citep{kingma2014adam} with a learning rate of $10^{-4}$ and a step-based scheduler with a decay gamma of $0.1$ for all model training. All disease classification and reconstruction models are trained for $50$ epochs, and the severity classification model is fine-tuned for $20$ epochs. For the policy model, we train it for $30$ epochs. Our experimental setup uses the PyTorch framework, and all computations are performed on an NVIDIA A100 Tensor Core GPU. Our code is publicly available at \url{https://github.com/vios-s/MRI_Sequential_Active_Sampling}.

\begin{table*}[htbp]
\centering
\caption{Average balanced accuracy (\%) and Area Under Curve (\%) results and the standard deviations of different diagnostic policies across varying sampling rates for ACL Sprain Diagnosis. Dashes (---) indicate scenarios where the method is not applicable. Bold numbers denote the best performance }
\label{tab:diagnostic_performance_acl}
\resizebox{\textwidth}{!}{
\begin{tabular}{|c| c c c c c|c c c c c| c c c c c|}  
\hline
\multicolumn{16}{|c|}{Balanced Accuracy $\uparrow$} \\
\hline
Diagnosis Task & \multicolumn{5}{c|}{Disease} & \multicolumn{5}{c|}{Severity Degree} & \multicolumn{5}{c|}{Sequential Diagnosis} \\
\hline
\textbf{Fully Sampled (Oracle)} &  &  & $\textbf{84.27}_{0.0}$ &  &  &  &  & $\textbf{69.20}_{1.5}$ &  &  &  &  & $\textbf{65.94}_{0.1}$ &  &  \\
\hline
Sampling Rate & 5\% & 9\%& 13\%& 17\%& 25\%& 5\% & 9\%& 13\%& 17\%& 25\% & 5\% & 9\%& 13\%& 17\%& 25\% \\
\hline
Undersampled & $72.90_{2.9}$&	$75.68_{2.5}$&	$76.14_{1.9}$&	$76.58_{1.8}$&	$79.64_{0.8}$& $\textbf{60.62}_{1.7}$&	$61.20_{3.0}$&	$62.90_{3.1}$&	$64.58_{4.0}$&	$68.42_{2.5}$&--- &--- &--- &--- &--- \\
Policy Classifier~\citep{du2024mri} & $72.90_{2.9}$ & $81.32_{1.7}$ & $81.78_{1.4}$ & $81.84_{0.7}$ & $82.94_{0.8}$ & $\textbf{60.62}_{1.7}$ & $71.94_{2.7}$ & $74.08_{2.3}$ & $\textbf{74.64}_{1.9}$ & $\textbf{78.58}_{2.6}$ &--- &--- &--- &--- &--- \\
Policy Recon~\citep{bakker2020experimental} &$62.54_{0.1}$&	$69.60_{0.4}$&	$78.60_{0.4}$&	$79.14_{0.2}$&	$81.92_{0.5}$&$37.70_{0.0}$&	$30.32_{1.9}$&	$43.08_{2.7}$&	$49.14_{0.1}$&	$54.64_{2.1}$&$41.90_{0.0}$&	$39.68_{0.7}$&	$52.44_{1.7}$&	$55.22_{0.1}$&	$60.40_{1.1}$  \\
Random Policy& $73.26_{2.8}$&	$75.02_{2.6}$&	$76.54_{2.9}$&	$77.48_{1.9}$&	$79.30_{1.2}$&	$58.94_{4.1}$&	$62.90_{3.2}$&	$61.46_{2.2}$&	$64.68_{1.8}$&	$66.32_{2.0}$&	$53.58_{3.1}$&	$56.00_{3.4}$&	$56.32_{1.8}$&	$59.64_{1.9}$&	$61.80_{1.4}$\\
\textbf{Weighted Policy (Ours)} & $\textbf{73.26}_{2.8}$ & $\textbf{84.20}_{1.3}$ & $\textbf{84.76}_{1.0}$ & $\textbf{84.34}_{0.7}$ & $\textbf{83.54}_{0.4}$ & $58.94_{4.1}$ & $\textbf{76.86}_{2.8}$ & $\textbf{74.10}_{3.1}$ & $73.88_{1.9}$ & $75.36_{2.8}$ & $\textbf{53.58}_{3.1}$ & $\textbf{73.42}_{2.0}$ & $\textbf{72.36}_{1.9}$ & $\textbf{71.72}_{1.7}$ & $\textbf{71.68}_{1.6}$ \\
\hline
\multicolumn{16}{|c|}{Area Under Curve $\uparrow$} \\
\hline
Diagnosis Task & \multicolumn{5}{c|}{Disease} & \multicolumn{5}{c|}{Severity Degree} & \multicolumn{5}{c|}{Sequential Diagnosis} \\
\hline
\textbf{Fully Sampled (Oracle)} &  &  & $\textbf{90.23}_{0.0}$ &  &  &  &  & $\textbf{71.03}_{1.2}$ &  &  &  &  & $\textbf{80.76}_{0.1}$ &  &  \\
\hline
Sampling Rate & 5\% & 9\%& 13\%& 17\%& 25\%& 5\% & 9\%& 13\%& 17\%& 25\% & 5\% & 9\%& 13\%& 17\%& 25\% \\
\hline
Undersampled & $\textbf{84.10}_{1.8}$&	$85.58_{1.3}$&	$86.50_{1.2}$&	$87.28_{1.3}$&	$88.40_{0.7}$& $\textbf{69.20}_{4.6}$&	$70.34_{5.5}$&	$73.38_{4.5}$&	$75.36_{4.2}$&	$79.70_{4.7}$&--- &--- &--- &--- &--- \\
Policy Classifier~\citep{du2024mri} & $\textbf{84.10}_{1.8}$ & $88.72_{0.5}$ & $88.64_{0.5}$ & $88.54_{0.4}$ & $88.98_{0.4}$ & $\textbf{69.20}_{4.6}$ & $83.78_{1.7}$ & $84.80_{1.3}$ & $85.10_{1.4}$ & $87.16_{1.5}$ &--- &--- &--- &--- &--- \\
Policy Recon~\citep{bakker2020experimental} &$78.16_{0.1}$&	$84.02_{0.1}$&	$87.98_{0.0}$&	$88.44_{0.1}$&	$89.30_{0.1}$& $35.80_{0.0}$&	$25.40_{0.5}$&	$42.10_{0.9}$&	$46.66_{0.4}$&	$57.86_{0.9}$&$56.98_{0.0}$&	$54.72_{0.3}$&	$65.02_{0.4}$&	$67.56_{0.2}$&	$73.58_{0.4}$\\
Random Policy& $83.24_{1.5}$&	$84.98_{0.7}$&	$85.98_{0.9}$&	$86.72_{0.8}$&	$87.98_{0.7}$&	$67.22_{5.5}$&	$72.06_{4.6}$&	$73.16_{3.8}$&	$74.52_{3.2}$&	$77.02_{2.7}$&	$75.22_{3.0}$&	$78.52_{2.2}$&	$79.54_{1.8}$&	$80.62_{1.8}$&	$82.50_{1.6}$\\
\textbf{Weighted Policy (Ours)} & $83.24_{1.5}$ & $\textbf{89.32}_{0.4}$ & $\textbf{89.54}_{0.2}$ & $\textbf{89.50}_{0.1}$ & $\textbf{89.36}_{\textbf{0.3}}$ & $67.22_{5.5}$ & $\textbf{88.14}_{1.2}$ & $\textbf{86.76}_{1.4}$ & $\textbf{86.08}_{1.0}$ & $\textbf{87.44}_{1.1}$ & $\textbf{75.22}_{3.0}$ & $\textbf{88.72}_{0.6}$ & $\textbf{88.14}_{0.8}$ & $\textbf{87.78}_{0.5}$ & $\textbf{88.40}_{0.7}$ \\
\hline
\end{tabular}
}

\centering
\caption{Same as in Table~\ref{tab:diagnostic_performance_acl} but for Cartilage Thickness Loss}
\label{tab:diagnostic_performance_cart}
\resizebox{\textwidth}{!}{
\begin{tabular}{|c| c c c c c|c c c c c| c c c c c|}  
\hline
\multicolumn{16}{|c|}{Balanced Accuracy $\uparrow$} \\
\hline
Diagnosis Task & \multicolumn{5}{c|}{Disease} & \multicolumn{5}{c|}{Severity Degree} & \multicolumn{5}{c|}{Sequential Diagnosis} \\
\hline
\textbf{Fully Sampled (Oracle)} &  &  & $\textbf{75.05}_{0.1}$ &  &  &  & & $\textbf{60.90}_{2.7}$ &  &  &  &  & $\textbf{54.84}_{0.0}$ &  &  \\
\hline
Sampling Rate & 5\% & 9\%& 13\%& 17\%& 25\%& 5\% & 9\%& 13\%& 17\%& 25\% & 5\% & 9\%& 13\%& 17\%& 25\% \\
\hline
Undersampled &$71.12_{3.0}$&	$73.22_{1.2}$&	$74.32_{0.9}$&	$74.62_{1.4}$&	$75.22_{1.4}$& $54.90_{2.3}$&	$59.42_{3.0}$&	$59.02_{3.0}$&	$60.32_{3.3}$&	$61.38_{2.8}$&--- &--- &--- &--- &--- \\
Policy Classifier~\citep{du2024mri} & $71.12_{3.0}$&	$73.72_{2.5}$&	$\textbf{75.14}_{1.1}$&	$\textbf{75.30}_{1.3}$&	$75.98_{1.2}$& $54.90_{2.3}$&	$58.58_{4.0}$&	$60.30_{1.4}$&	$\textbf{63.10}_{3.9}$&	$\textbf{63.84}_{5.0}$&--- &--- &--- &--- &--- \\
Policy Recon~\citep{bakker2020experimental} &$64.34_{0.1}$&	$65.86_{0.9}$&	$72.18_{0.3}$&	$71.20_{0.3}$&	$75.92_{0.4}$&$56.58_{0.0}$&	$56.68_{1.0}$&	$54.50_{0.9}$&	$55.18_{0.0}$&	$56.60_{0.9}$&$45.22_{0.0}$&	$44.72_{1.3}$&	$50.18_{0.7}$&	$50.16_{0.1}$&	$54.46_{0.7}$\\
Random Policy& $72.76_{1.5}$&	$\textbf{73.88}_{1.6}$&	$75.08_{0.8}$&	$75.20_{1.3}$&	$\textbf{76.06}_{1.8}$&	$\textbf{58.92}_{5.7}$&	$\textbf{61.66}_{6.8}$&	$\textbf{61.60}_{4.8}$&	$62.22_{3.9}$&	$63.64_{3.9}$&	$\textbf{54.40}_{2.8}$&	$54.66_{4.2}$&	$\textbf{55.64}_{1.9}$&	$\textbf{55.80}_{1.9}$&	$\textbf{58.44}_{2.9}$\\
\textbf{Weighted Policy (Ours}) &$\textbf{72.76}_{1.5}$&	$72.96_{1.0}$&	$74.02_{0.9}$&	$73.86_{1.1}$&	$73.84_{1.5}$&$\textbf{58.92}_{5.7}$&	$60.62_{3.9}$&	$58.98_{2.8}$&	$60.88_{2.5}$&	$62.32_{3.0}$&$\textbf{54.40}_{2.8}$&	$\textbf{54.88}_{2.1}$&	$53.90_{2.2}$&	$54.80_{0.9}$&	$55.48_{0.6}$ \\
\hline
\multicolumn{16}{|c|}{Area Under Curve $\uparrow$} \\
\hline
Diagnosis Task & \multicolumn{5}{c|}{Disease} & \multicolumn{5}{c|}{Severity Degree} & \multicolumn{5}{c|}{Sequential Diagnosis} \\
\hline
\textbf{Fully Sampled (Oracle)} & & & $\textbf{88.23}_{0.0}$ & &  &  &  & $\textbf{73.58}_{4.2}$ &  &  &  &  & $\textbf{80.88}_{0.5}$ &  &  \\
\hline
Sampling Rate & 5\% & 9\%& 13\%& 17\%& 25\%& 5\% & 9\%& 13\%& 17\%& 25\% & 5\% & 9\%& 13\%& 17\%& 25\% \\
\hline
Undersampled & $84.58_{1.0}$&	$86.02_{0.7}$&	$87.38_{0.3}$&	$87.86_{0.7}$&	$88.08_{0.6}$& $61.90_{5.0}$&	$66.40_{7.4}$&	$68.76_{5.3}$&	$71.80_{5.4}$&	$76.82_{4.5}$&--- &--- &--- &--- &--- \\
Policy Classifier~\citep{du2024mri}&$84.58_{1.0}$&	$86.74_{0.4}$&	$87.32_{0.4}$&	$87.82_{0.4}$&	$87.96_{0.4}$&$61.90_{5.0}$&	$\textbf{81.12}_{2.0}$&	$\textbf{83.10}_{1.6}$&	$83.24_{1.3}$&	$85.26_{1.4}$&--- &--- &--- &--- &--- \\

Policy Recon~\citep{bakker2020experimental}&$83.72_{0.0}$&	$85.30_{0.1}$&	$86.18_{0.1}$&	$86.42_{0.0}$&	$87.38_{0.1}$&$\textbf{71.44}_{0.1}$&	$68.32_{0.9}$&	$71.30_{0.6}$&	$70.52_{0.6}$&	$75.00_{0.9}$&$\textbf{77.58}_{0.0}$&	$76.84_{0.4}$&	$78.74_{0.3}$&	$78.48_{0.3}$&	$81.22_{0.4}$\\
Random Policy& $85.24_{0.8}$&	$86.18_{1.0}$&	$86.98_{0.3}$&	$87.52_{0.6}$&	$88.44_{0.9}$&	$65.72_{6.1}$&	$70.28_{5.0}$&	$71.04_{5.0}$&	$75.54_{4.7}$&	$78.52_{2.9}$&	$75.46_{3.2}$&	$78.20_{2.7}$&	$79.00_{2.6}$&	$81.54_{2.4}$&	$83.48_{1.5}$\\
\textbf{Weighted Policy (Ours)} &$\textbf{85.24}_{0.8}$&	$\textbf{87.34}_{0.3}$&	$\textbf{89.06}_{0.4}$&	$\textbf{88.98}_{0.3}$&	$\textbf{89.02}_{0.3}$&$65.70_{6.1}$&	$75.04_{1.2}$&	$80.94_{1.6}$&	$\textbf{83.60}_{1.0}$&	$\textbf{85.58}_{1.5}$&$75.44_{3.2}$&	$\textbf{81.20}_{0.6}$&	$\textbf{84.98}_{0.8}$&	$\textbf{86.28}_{0.6}$&	$\textbf{87.26}_{0.8}$  \\
\hline
\end{tabular}
}
\end{table*}

\section{Experimental Results and Analysis}
We divide this section into several parts, each addressing key questions or hypotheses. First, we aim to show that our approach maintains diagnostic performance when compared to other non-sequential and single objective policies. We then proceed to address our main hypothesis that adopting sequential decisions leads to $k$-space savings. Finally, we show that the natural ordering of the objectives results in even more savings.

\subsection{Comparison to Non-sequential and Single Objective Policies} 
In this section, we first answer a critical question: \textit{Will our method compromise performance when enforcing multiple objectives in a single scan?} We compare the performance of our sequential diagnosis method with baselines and benchmarks designed for a single binary diagnosis task, namely the \textit{Undersampled},  \textit{Policy Recon}, and \textit{Policy Classifier}, at various sample rates for inferring the conditions of ACL Sprain and Cartilage Thickness Loss.
\textit{Undersampled} randomly samples $k$-space lines progressively. \textit{Policy Recon}, \textit{Policy Classifier}, and \textit{Weighted Policy} start with a randomly initialized mask at a sample rate of $5\%$ and perform sampling decision-making respectively using their policies (see Section~\ref{sec: baselines}). We also include \textit{Fully sampled} as a diagnostic oracle, which does not have a sampling budget. Additionally, \textit{Random Policy} is included as the direct baseline for \textit{Weighted Policy}, which provide random guidance. The results are shown in Table~\ref{tab:diagnostic_performance_acl} and Table~\ref{tab:diagnostic_performance_cart}.

Our \textit{Weighted Policy}, consistently demonstrates strong performance in all metrics across different sampling rates and both tasks (ACL Sprain and Thickness Loss). Focusing first on Table~\ref{tab:diagnostic_performance_acl}, we observe that: 
(i) \textit{Weighted Policy} approaches the performance of  \textit{Fully Sampled}, showing that careful selection of $k$-space lines avoids taking redundant and irrelevant information. 
(ii) The advantage of policy-based optimization in our method is clearly demonstrated as we can approach the performance of \textit{Fully Sampled} with roughly $10\%$ $k$-space data across all subtasks. 
(iii) As expected, \textit{Policy Recon}, although providing benchmark reconstruction performance as described in~\citep{bakker2020experimental}, underperforms our approach, which optimizes for each task at hand. 
(iv) More importantly, our \textit{Weighted Policy}, which jointly optimizes disease detection and severity quantification, outperforms the \textit{Policy Classifier} in most cases even when used for a single subtask with fewer samples. In fact, with $10\%$ sampling in disease detection ($84.20\%$ vs. $81.32\%$) and $20\%$ $k$-space sampling for severity quantification ($74.10\%$ vs. $74.08\%$), this finding is extremely important, as it answers the initially posed question. It demonstrates that a policy optimized for multiple objectives does not compromise diagnostic performance compared to policies optimized for a single objective.

Similar observations hold for Cartilage Thickness Loss in Table~\ref{tab:diagnostic_performance_cart}, evidently illustrating the applicability of our methods in two settings and hence the strength of our findings.

\begin{figure*}[t]
    \centering
    \includegraphics[width=\textwidth]{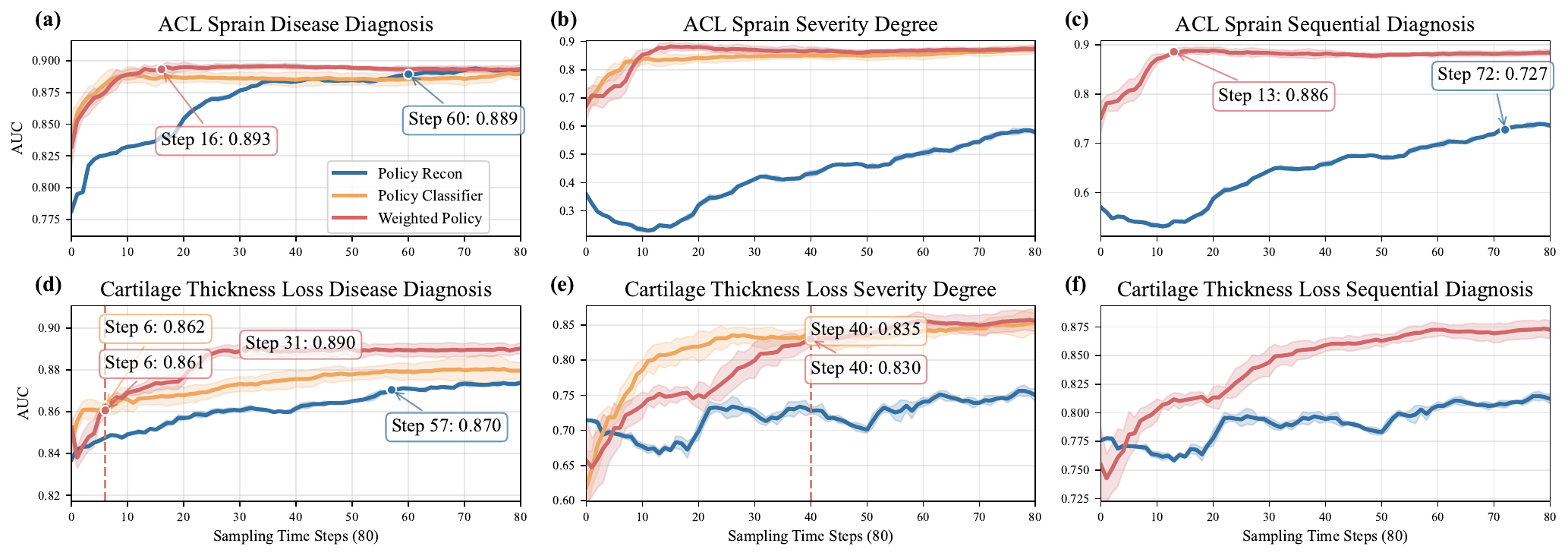}
    \caption{Performance comparison of our proposed \textit{Weighted Policy} and two single-task oriented policies. The horizontal axis indicates the cumulative lines acquired when sampling. The $80$-step active sampling procedure starts from $16$ randomly sampled $k$-space lines. Earlier plateau indicates fewer required $k$-space lines.} 
    \label{fig: kspace_saving}
\end{figure*}

\begin{figure}[t]
    \centering
    \includegraphics[width=1.0\linewidth]{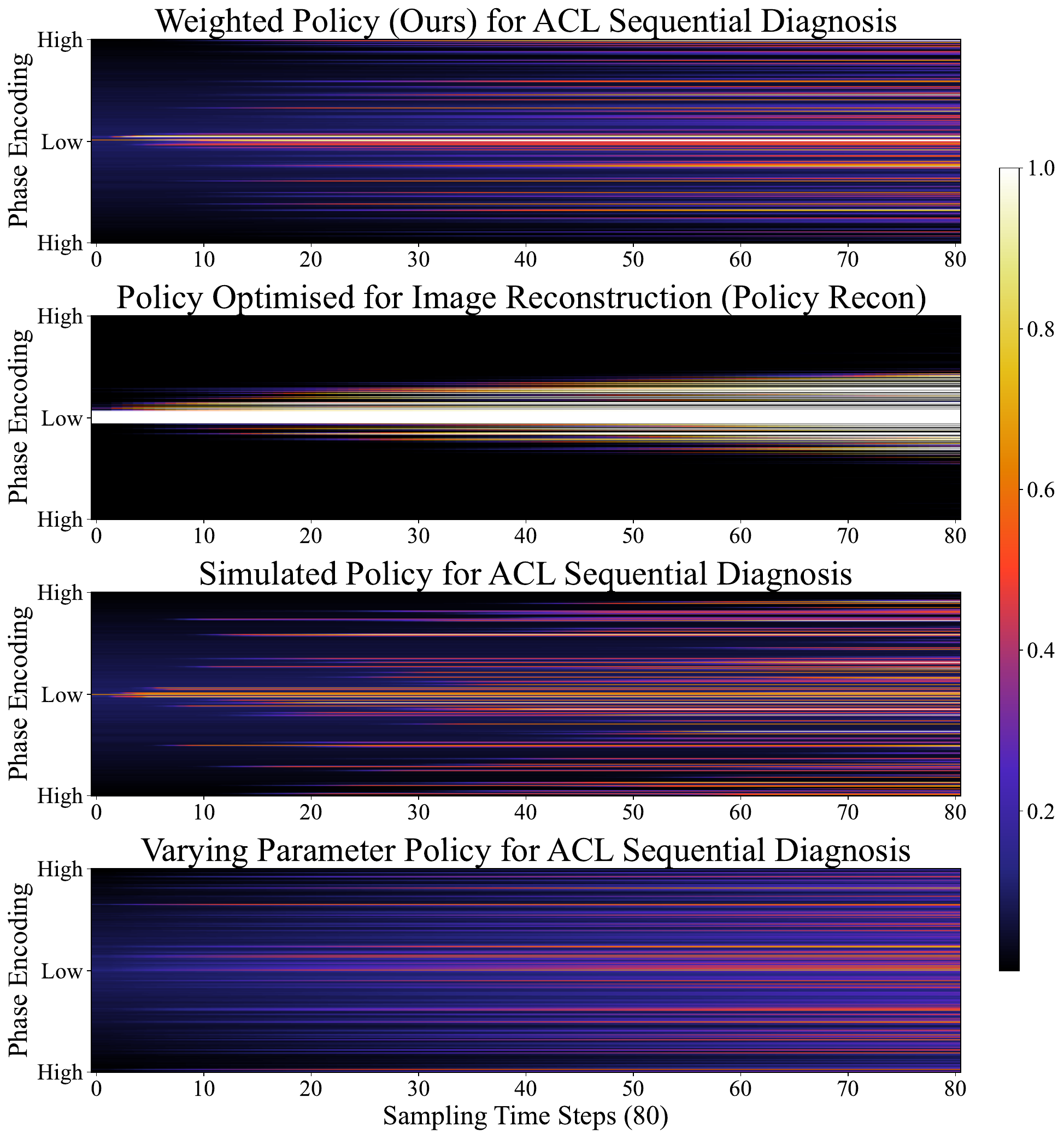}
    \caption{Average $k$-space sampling behaviors across all subjects of our proposed \textit{Weighted Policy} and benchmarks. A High value in the plot represents a high probability to be sampled. The horizontal axis indicates the cumulative lines acquired when sampling.}
    \label{fig: mask}
\end{figure}

\subsection{$k$-space Savings} 

We now proceed to answer our main hypothesis: \textit{Joint optimization of a policy for two objectives ultimately leads to sampling efficiency and savings}.

In Figure~\ref{fig: kspace_saving}, we illustrate the diagnostic performance of policies that progressively make decisions on which line to acquire, line-by-line, and visualize their sampling trajectories. We present the performance of our method and compare with \textit{Policy Recon} and \textit{Policy Classifier} to showcase the difference between sampling patterns of single-task methods versus ours.

We find that the performance curve plateaus during the sampling process, indicating that the policies might have collected the majority of diagnostically significant $k$-space lines that contribute to diagnostic performance enhancement.

For disease detection, as shown in Figure~\ref{fig: kspace_saving} (a) and (d), compared to \textit{Policy Recon} which optimizes sampling trajectories for reconstruction, our method plateaus earlier at step $16$ for ACL Sprain, whereas \textit{Policy Recon} plateaus much later (at step $60$ for ACL Sprain leading to considerable $k$-space savings. 
The advantage of a diagnosis-oriented policy is further highlighted in severity quantification (Figure~\ref{fig: kspace_saving} (b)), where our method successfully identifies $k$-space lines that can significantly improve inference. In contrast, although it can achieve continuous improvements in reconstruction performance, \textit{Policy Recon} struggles to effectively locate diagnostically relevant $k$-space lines and produce comparable outcomes in both severity and sequential diagnosis classification.
Consequently, when evaluating sequential diagnosis in ACL Sprain (Figure~\ref{fig: kspace_saving} (c)), our method reaches a plateau at step $13$ with a high AUC of $0.886$, while \textit{Policy Recon} arrives at step $72$ with the AUC of $0.727$. Similar observations hold for Cartilage Thickness Loss (Figure~\ref{fig: kspace_saving} (d) to (f)).

In Figure~\ref{fig: kspace_saving}, compared to the policies of \textit{Policy Classifier} that optimize sampling trajectories independently for separate tasks (disease detection or severity quantification), our method provides competitive results and achieves considerable acquisition savings. For ACL Sprain, our policy quickly approaches the performance of \textit{Policy Classifier}s, and exceeds them afterward. Similarly, as shown in plots (d) and (e), \textit{Weighted Policy} samples $6$ lines in disease detection and $40$ lines in severity quantification for Cartilage Thickness Loss to achieve the performance of the corresponding \textit{Policy Classifier}. Notably, our policy achieves this sequentially whereas the other policies require different sessions for two tasks. Hence, our method can significantly reduce the quantity of necessary $k$-space lines and thereby shorten the acquisition time.

\begin{table}
\centering
\caption{Sampling and Computational Efficiency Metrics evaluated on a single NVIDIA A100 Tensor Core GPU. n/a refers to non applicable.}
\label{tab:efficiency_metrics}
\resizebox{\linewidth}{!}{
\begin{tabular}{|c|c|c|c|}
\hline
\textbf{Method}  & \textbf{Policy Classifier}~\citep{du2024mri}& \textbf{Policy Recon}~\citep{bakker2020experimental}& \textbf{Weighted Policy (Ours)} \\
\hline
\multicolumn{4}{|c|}{\textit{Sampling Efficiency}} \\
\hline
Initial 5\% Acquisition (ms) & 20.00 &20.00 &20.00\\
Selection Time (ms/step) &14.33& 29.67 & 26.53 \\
\hline
\multicolumn{4}{|c|}{\textit{Computational Efficiency}} \\
\hline
Inference Time (ms) &12.62& 29.93 & 25.31 \\
Reconstruction Time (ms) &n/a& 3.73 & n/a \\
Total Pipeline Latency (ms) &12.62& 33.66 & 25.31 \\
Peak Memory Usage (GB) & 0.58& 1.19 & 1.11 \\
\hline
\end{tabular}}
\end{table}

\begin{figure*}[t]
\begin{center}
\includegraphics[width=1.0\textwidth]{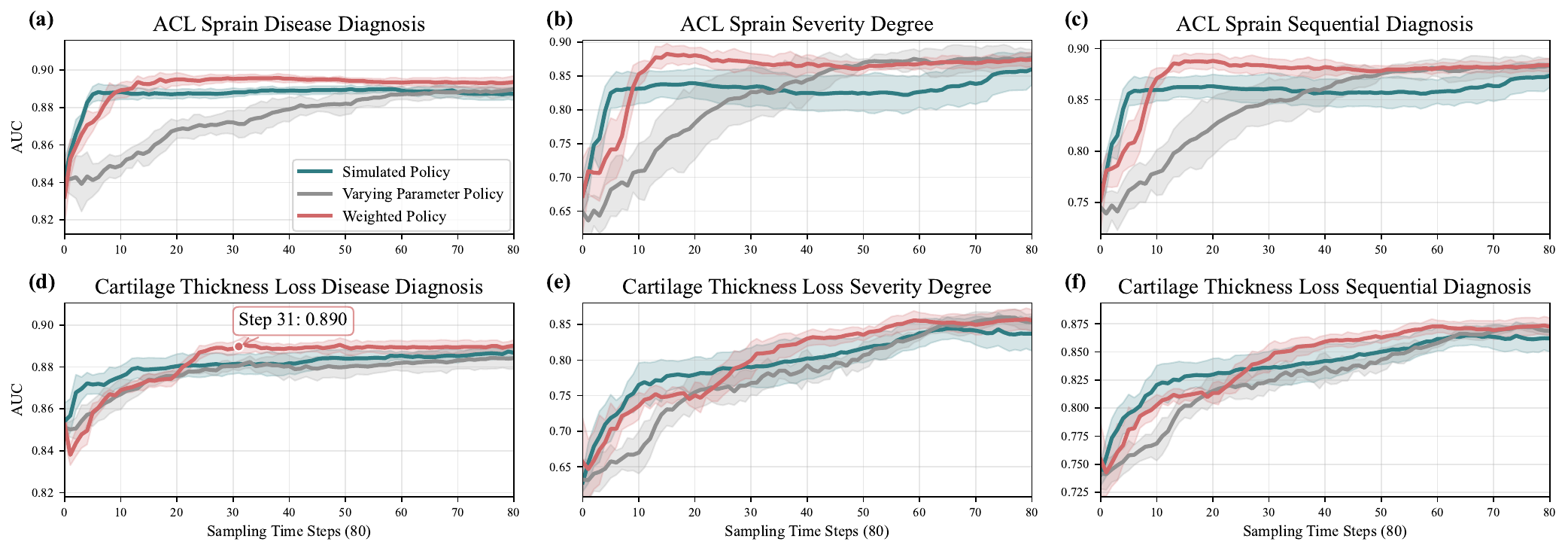}
\end{center}
\caption{Performance comparison of our proposed \textit{Weighted Policy} and two multi-task oriented policies. The horizontal axis indicates the cumulative lines acquired when sampling. The $80$-step active sampling procedure starts from $16$ randomly sampled $k$-space lines. Earlier plateau indicates fewer required $k$-space lines.} \label{fig: MORL}
\end{figure*}

Sampling efficiency and inference latency are critical for clinical viability. As reported in Table~\ref{tab:efficiency_metrics}, the initial $5\%$ acquisition requires $20$ ms, and the policy requires $26.53$ ms per subsequent $k$-space line selection for the proposed method and $4.62$ ms more for \textit{Policy Recon}. In ACL sprain sequential diagnosis, performance plateaus at step $13$ for the proposed method versus step $72$ for \textit{Policy Recon}, corresponding to a $1791.35$ ms reduction in total sampling time. Compared to the \textit{Policy Recon}, our method achieves additional savings by eliminating reconstruction ($3.73$ ms) and performing direct $k$-space inference ($29.93$ ms versus $25.31$ ms). Notably, if we consider a ``trigger" system with two separate \textit{Policy Classifier}s for two objectives, which requires re-initiating the sampling protocol upon positive disease identification, the system incurs supplementary overhead comprising initial $5\%$ acquisition time ($20$ ms) and extended $k$-space line selection ($33.29$ ms). Collectively, such computational efficiencies reduce acquisition time without compromising diagnostic accuracy,and translate into even more substantial time savings during practical MRI acquisition. Consider a typical Fast Spin Echo (FSE) sequence with a repetition time (TR) of $2000$ ms and an Echo Train Length of $4$, encoding four $k$-space lines per TR. Following a universal $4$-TR initial sampling phase, existing approaches require significantly more scan time: A ``trigger" system including two \textit{Policy Classifier}s needs $25$ lines ($15$ for disease identification and $10$ for severity quantification), requiring $15$ TRs (including re-initiation) for $30,000$ ms, while the \textit{Policy Recon} baseline requires $72$ lines, necessitating $22$ TRs for $44,000$ ms. In contrast, our method achieves its diagnostic goal in just $13$ steps, requiring only $8 $TRs and $16,000$ ms of scan time. This yields time savings of $14,000$ ms compared to the \textit{Policy Classifier} and $28,000$ ms compared to \textit{Policy Recon}, demonstrating how our method reduces the actual time patients spend in the scanner—a critical advantage for clinical viability.

Additionally, we want to compare the sampling behaviors of different policies averaged across the subject population. As shown in Figure~\ref{fig: mask}, referring to the top two plots, there are evident differences between the trajectories of \textit{Policy Recon} and \textit{Weighted Policy}. 
\textit{Policy Recon} tends to consistently sample low-frequency central $k$-space lines, whereas our method shows a broader diversity (across the frequency spectrum), illustrating its adaptation at the patient level. 

\subsection{Sequential Diagnosis in One Scanning Session}

We base our work on a key question: \textit{Is it necessary to adopt a sequential process, detecting disease first and then assessing severity?} This question is crucial because our proposed sequential diagnosis workflow employs the lexicographic ordering in step-wise weighting module, attempting to mirror the clinical workflow of radiologists. To address this key question, we train two policies with different ways to optimize for multiple objectives in parallel and independently. 
The first, termed \textit{Simulated Policy}, uses the same architecture as ours but employs a single-objective RL algorithm rewarded by the performance of a three-way diagnosis (no finding, low, or high severity) without specifying the lexicographic ordering among diagnostic objectives. The second, the \textit{Varying Parameter Policy}, is instead a MORL-based method, which uses separate policies to optimize detection and quantification in parallel as independent objectives, and the final sampling policy is the average of the result of the two policy networks. 

Referring to Figure~\ref{fig: MORL} and first focusing on comparisons with the \textit{Simulated Policy}, we see that the \textit{Simulated Policy} plateaus earlier at a lower performance compared to our method. 
This is because the \textit{Simulated Policy} is rewarded based on overall sequential diagnosis improvements, forcing the policy to find $k$-space lines that contribute to a three-way objective, therefore lacking reward information for each independent objective. Our method explicitly calculates the separate rewards of the two objectives, which simplifies the sequential diagnosis task and reduces the complexity of the sampling strategy optimization.  

We compare with \textit{Varying Parameter Policy}, a MORL-based trained policy, whose key differences are: (i) it employs individual policies to process rewards from the independent objectives in parallel; and most importantly, (ii) it does not utilize the lexicographic ordering between objectives. Thus, we expect to observe worse performance for this `simpler' policy. 
Looking again at Figure~\ref{fig: MORL}, and now comparing the `grey' (\textit{Varying Parameter Policy}) and `red' (\textit{Weighted Policy}) lines, our method outperforms the other policy and plateaus earlier at a higher accuracy level, particularly in the overall sequential diagnosis (Figure~\ref{fig: MORL} (c) and (f)), hence achieving $k$-space sampling savings and in turn less scanning time.
This highlights the benefits of lexicographic ordering in multi-objective optimization and step-wise weighting.
Our policy thus learns the inherent shared diagnostic features between disease detection and severity quantification, and is encouraged to seek mutually beneficial lines during sampling.   
Consequently, in Figure~\ref{fig: MORL} Cartilage Thickness Loss, after sampling what is needed to decide in the presence of disease (shown as plateauing at step $31$ subplot (d), the optimization prioritizes sampling of $k$-space lines that contribute to severity as observed by an increase in performance in the subplot (e) without compromising disease diagnosis performance (the plateau). 

We can observe such transitions also at the level of sampling trajectories in Figure~\ref{fig: mask} as well. Looking at the bottom two plots and comparing them with the topmost, until sampling step $16$, our proposed policy samples more low-frequency lines (containing structural information) that can contribute to both objectives and patient-level disease-related information (shown as purple in the plot). Then, the policy smoothly shifts to sampling diagnostic features located in high-frequency zone, which may contribute to severity quantification for ACL Sprain. While \textit{Varying Parameter Policy} tends to find patient-level features from the start, it can miss the mutual diagnostically significant lines between disease detection and severity quantification. The probability trajectories of the \textit{Simulated Policy} show several frequency locations with high sampling probability, illustrating a tendency to generate similar sampling trajectories for all subjects, thereby lacking exploration of patient-level diagnostic information.

\begin{table*}[t]
\centering
\caption{Performance comparison of \textit{Weighted Policy} (ACL Sprain) against
\textit{Fully Sampled (Oracle)} baseline across different sampling rates. $\Delta$ is the difference of means from the Oracle, in percentage points for rates and in cases for High Grade Missed; disease-positive test subjects $n=107$, of whom $65$ high grade. The initial $5\%$ samples are randomly
sampled. All values are means over $3$ random seeds, with standard
deviation across seeds given as a subscript. Disease Miss Rate and High Grade Missed are
computed over all disease-positive subjects; Severity Overestimation and Underestimation
are computed only among subjects correctly detected as diseased (Section~\ref{sec:eval_metrics}).}
\label{tab:diagnostic_risk_results}
\resizebox{\textwidth}{!}{
\begin{tabular}{|c|c|c|c|c|}
\hline
\textbf{Sampling Rate} & \textbf{Disease Miss Rate (\%)}$\downarrow$ & \textbf{High Grade Missed (n)}$\downarrow$ & \textbf{Severity Overestimation (\%)}$\downarrow$ & \textbf{Severity Underestimation (\%)}$\downarrow$\\
\hline
\multicolumn{5}{|c|}{\textit{Fully sampled (Oracle)}} \\
\hline
100\% & $14.02_{0.00}$ & $7.0_{0.0}$ & $20.59_{0.00}$ & $43.10_{0.00}$ \\
\hline
\multicolumn{5}{|c|}{\textit{Weighted Policy (Ours)}} \\
\hline
5\% & $33.64_{14.14}$ (+19.6) & $25.0_{13.1}$ (+18.0) & $20.43_{4.39}$ (-0.2) & $48.33_{32.96}$ (+5.2) \\
9\% & $17.13_{1.08}$ (+3.1) & $12.7_{0.6}$ (+5.7) & $11.01_{8.07}$ (-9.6) & $29.94_{12.97}$ (-13.2) \\
13\% & $15.89_{2.47}$ (+1.9) & $11.3_{2.1}$ (+4.3) & $11.93_{7.01}$ (-8.7) & $26.71_{15.27}$ (-16.4) \\
17\% & $17.13_{0.54}$ (+3.1) & $12.7_{1.2}$ (+5.7) & $11.93_{8.45}$ (-8.7) & $25.48_{16.39}$ (-17.6) \\
25\% & $18.38_{0.54}$ (+4.4) & $13.7_{0.6}$ (+6.7) & $10.19_{8.02}$ (-10.4) & $25.97_{16.03}$ (-17.1) \\
\hline
\end{tabular}}
\end{table*}

\subsection{Diagnostic Risk and Safety Evaluation}
In this section, we investigate the diagnostic risk and safety of \textbf{Weighted Policy} for ACL sprain diagnosis, in comparison with \textbf{Fully Sampled (Oracle)} as summarized in Table~\ref{tab:diagnostic_risk_results}.
The critical clinical risk is missed detection (false negatives), which delays diagnosis and appropriate treatment~\citep{arastu2015prevalence}. The disease miss rate demonstrates that our method can effectively approach the performance of the \textit{Fully Sampled} baseline with only $10$\% sampling, achieving a disease miss rate of $15.73$\% compared to $14.04$\% for the oracle ($+1.7$ \%). 
We also evaluate how many patients with high-grade ACL tears are missed during detection. At $10$\% sampling, our method misses $11.6$ high-grade cases compared to $7$ for \textit{Fully Sampled}, representing an increase of $4.6$ cases. This modest increase must be interpreted in the context of reduced acquisition time and may have different implications depending on the clinical setting.

Regarding severity assessment errors, our \textbf{Weighted Policy} exhibits surprising characteristics. Severity overestimation rate decreases substantially compared to the oracle baseline across all sampling rates, with the most pronounced reduction at $10$\% sampling ($4.89$\% vs. $20.59$\%, $\Delta = -15.7$). Similarly, Weighted Policy outperforms oracle performance in Severity underestimation rate at 10\% sampling ($35.71$\%, $\Delta = -7.2$). The superiority in severity estimation proving its diagnostic safety while enabling substantial savings in medical resources and appropriate healthcare to each patient.

Overall, the \textbf{Weighted Policy} at $10$\% sampling demonstrates a positive performance-efficiency trade-off, maintaining disease detection performance comparable to \textit{Fully Sampled} while substantially reducing acquisition time. The observed increase in high-grade missed cases ($4.6$ additional cases) should be interpreted in the context of faster scanning, which contributes to improved patient comfort, reduced motion artifacts, and increased throughput.

\subsection{Reconstruction Compatibility of Task-specific Sampling}
\label{sec: recon_compatibility}

\begin{figure*}
\centering
\includegraphics[width=0.9\textwidth]{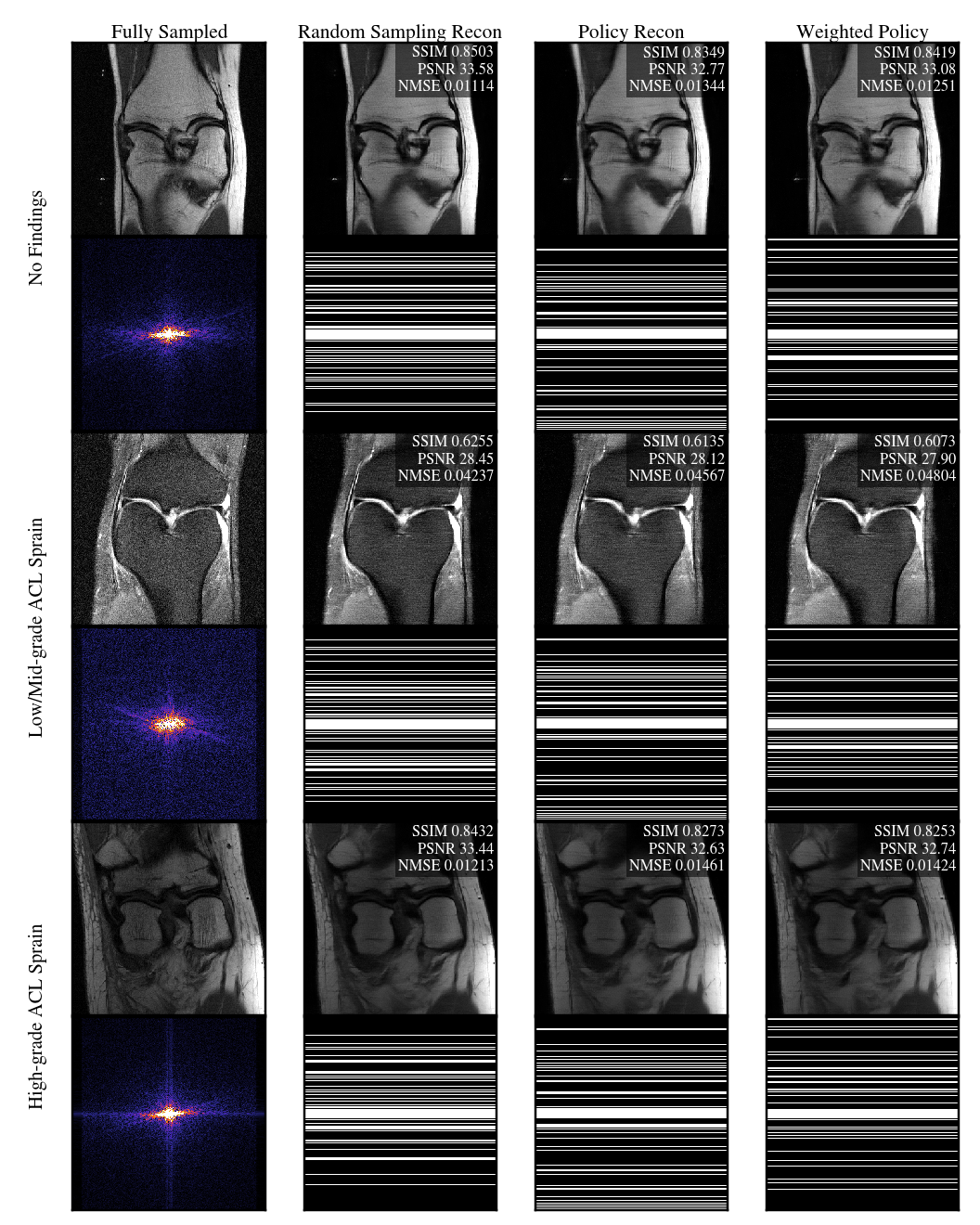}
\caption{Reconstruction compatibility of sequential task-specific sampling for ACL Sprain at a $5\%$ centre fraction. Rows correspond to no findings, low-moderate grade, and high-grade ACL Sprain. The first column shows the fully sampled reference and its $k$-space; the remaining columns show reconstructions of the final undersampled $k$-space obtained by random sampling, \textit{Policy Recon}, and the \textit{Weighted Policy}, all reconstructed with the same U-Net trained under conventional random sampling. Per-example SSIM, PSNR, and NMSE are annotated. Random sampling is the in-distribution reconstruction reference.}
\label{fig: recon_acl}
\end{figure*}

\begin{figure*}
\centering
\includegraphics[width=0.9\textwidth]{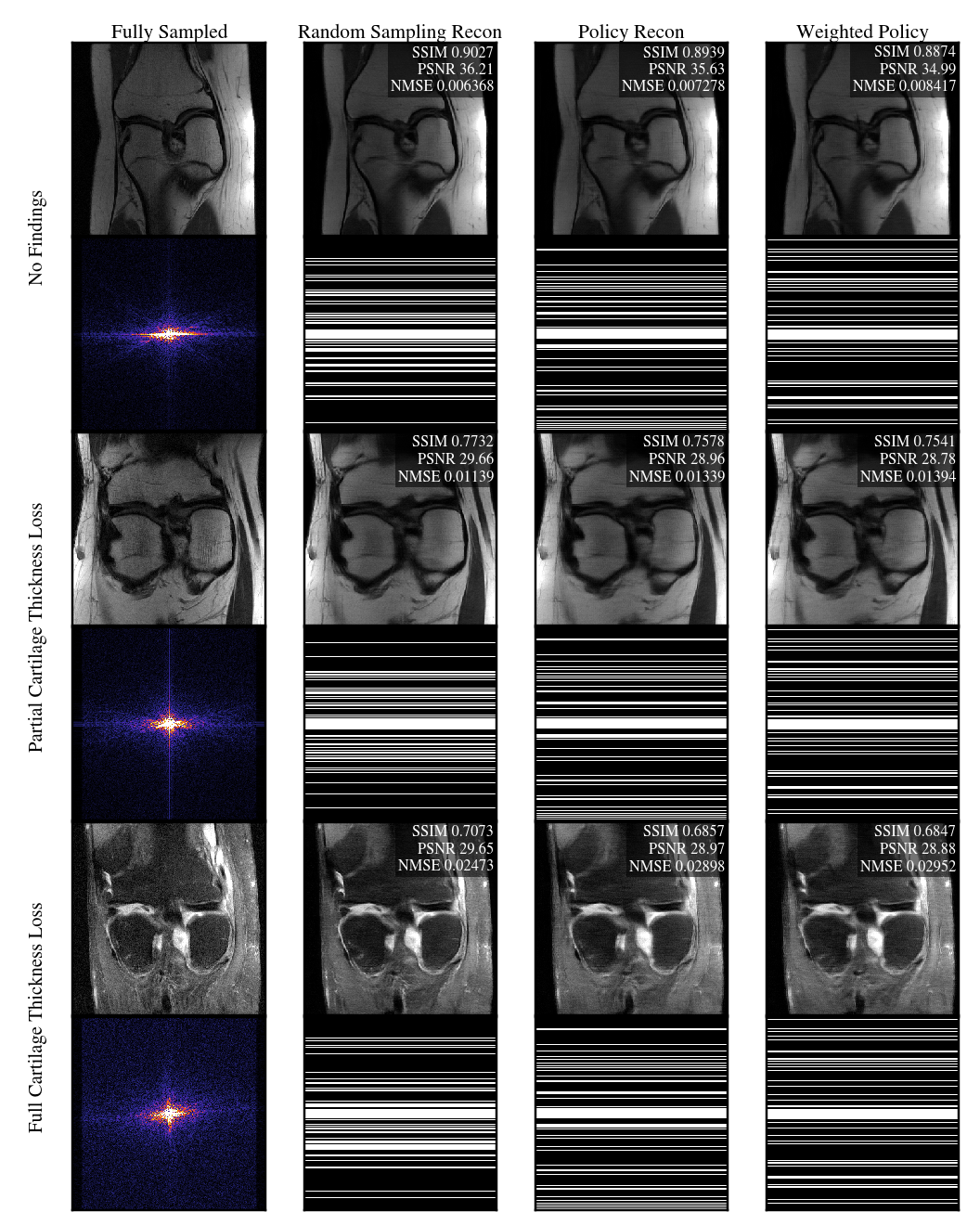}
\caption{Reconstruction compatibility of sequential task-specific sampling for Cartilage Thickness Loss at a $5\%$ centre fraction. Rows correspond to no findings, partial, and full Cartilage Thickness Loss. The first column shows the fully sampled reference and its $k$-space; the remaining columns show reconstructions of the final undersampled $k$-space obtained by random sampling, \textit{Policy Recon}, and the \textit{Weighted Policy}, all reconstructed with the same U-Net trained under conventional random sampling. Per-example SSIM, PSNR, and NMSE are annotated. Random sampling is the in-distribution reconstruction reference.}
\label{fig: recon_cart}
\end{figure*}

Because the \textit{Weighted Policy} selects $k$-space by sequential diagnostic value instead of image fidelity, the question examined for single-task sampling in Chapter~4 applies here as well: whether the acquired measurements can still be used for reconstruction. The final $k$-space obtained by random sampling, \textit{Policy Recon} and the \textit{Weighted Policy} is reconstructed by the same U-Net~\citep{ronneberger2015u}, trained under a conventional random-sampling protocol, at a $25\%$ sampling budget with a $5\%$ centre fraction, matching the setting of Chapter~4. Random sampling serves as the in-distribution reference, and \textit{Policy Recon}~\citep{bakker2020experimental} represents reconstruction-optimised acquisition. Table~\ref{tab:recon_aggregate} reports SSIM~\citep{wang2004image}, NMSE and PSNR over the whole test set, and Figures~\ref{fig: recon_acl} and~\ref{fig: recon_cart} show representative cases.

\begin{table}[t]
\centering
\caption{Reconstruction quality on the $1{,}841$ test slices at a $25\%$ sampling
budget with a $5\%$ centre fraction. SSIM, NMSE and PSNR are computed per slice
and averaged over slices; all rows are means over $3$ seeds, with standard
deviation across seeds given as a subscript. All undersampled $k$-space is
reconstructed by the same U-Net; the zero-filled row is the random mask without
reconstruction. ACL: ACL Sprain; CART: Cartilage Thickness Loss.}
\label{tab:recon_aggregate}
\resizebox{0.5\textwidth}{!}{
\begin{tabular}{|c|c|c|c|}
\hline
\textbf{Method} & \textbf{SSIM $\uparrow$} & \textbf{NMSE $\downarrow$} & \textbf{PSNR (dB) $\uparrow$} \\
\hline
Zero-filled input & $0.5987_{0.0002}$ & $0.0586_{0.0001}$ & $26.31_{0.00}$ \\
Random sampling Recon & $\mathbf{0.6681}_{0.0002}$ & $\mathbf{0.0329}_{0.0001}$ & $\mathbf{29.27}_{0.00}$ \\
\hline
Policy Recon~\citep{bakker2020experimental} & $0.6534_{0.0001}$ & $0.0352_{0.0000}$ & $28.76_{0.00}$ \\
Weighted Policy, ACL & $0.6527_{0.0002}$ & $0.0355_{0.0001}$ & $28.72_{0.00}$ \\
Weighted Policy, CART & $0.6548_{0.0001}$ & $0.0348_{0.0000}$ & $28.84_{0.00}$ \\
\hline
\end{tabular}
}
\end{table}

The \textit{Weighted Policy} produces reconstructions numerically similar to reconstruction-driven sampling. Relative to \textit{Policy Recon} (SSIM $0.6534$, PSNR $28.76$ dB), it differs by $-0.0007$ SSIM and $-0.04$ dB for ACL Sprain and by $+0.0014$ SSIM and $+0.08$ dB for Cartilage Thickness Loss. Against the corresponding single-objective policies of Chapter~4 (Table~\ref{tab:chapter4_reconstruction_aggregate}), evaluated on the same test slices with the same reconstructor, the differences are of similar size ($+0.0037$ SSIM and $-0.02$ dB for ACL Sprain; $-0.0042$ SSIM and $-0.16$ dB for Cartilage Thickness Loss). Both \textit{Weighted Policy} masks improve on the zero-filled input (SSIM $0.5987$, PSNR $26.31$ dB) by about $0.05$ in SSIM and $2.4$ to $2.5$ dB in PSNR.

Random sampling remains the strongest reconstruction condition, but only by a small margin. It gives the highest SSIM ($0.6681$), the lowest NMSE ($0.0329$) and the highest PSNR ($29.27$ dB), placing the \textit{Weighted Policy} $0.013$ to $0.015$ lower in SSIM and $0.43$ to $0.55$ dB lower in PSNR. As in Chapter~4, some gap is expected, since the reconstructor was trained on random masks, and its size indicates that sequential diagnostic acquisition retains most of the information the reconstructor relies on.

The representative cases in Figures~\ref{fig: recon_acl} and~\ref{fig: recon_cart} illustrate this at each severity level. The figures are organised by severity: no findings, low-moderate and high-grade ACL Sprain, and no findings, partial and full Cartilage Thickness Loss. At every severity level, the \textit{Weighted Policy} reconstructions preserve joint structures and tissue boundaries, even though part of the sampling budget is spent on higher-frequency lines selected for sequential diagnosis. The annotated per-example metrics are lower than for random sampling, consistent with the aggregate results. 

As in Chapter~4, $k$-space collected for sequential diagnosis remains usable by a conventional reconstructor, at a quality numerically similar to reconstruction-driven sampling and close to the in-distribution reference, so images can still be formed when visual review is required. The same caveats apply: the metrics measure image similarity only, radiologists did not assess the reconstructed images, and the reconstructor was not adapted to the structured masks.

\begin{figure*}[t]
\centering
\includegraphics[width=1.0\textwidth]{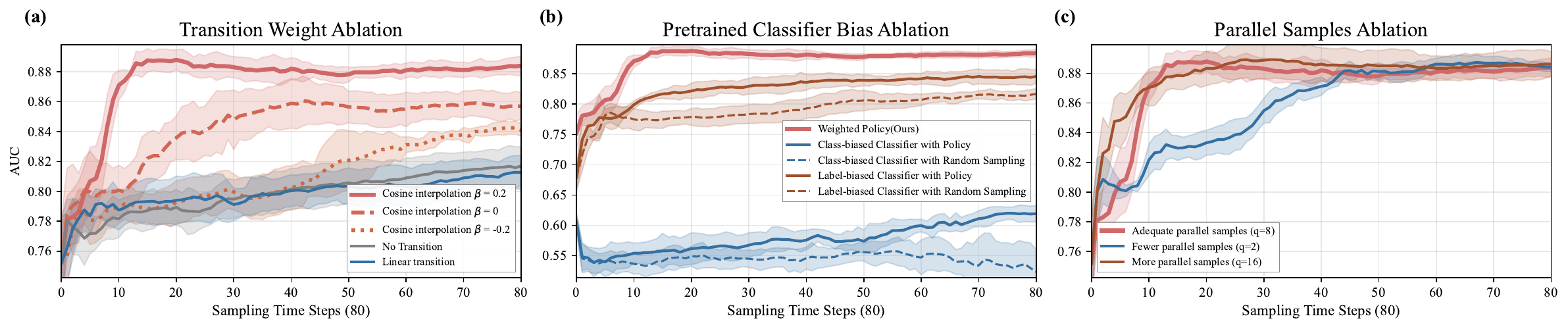}
\caption{Ablation studies on ACL sprain diagnosis. We evaluate the sensitivity of our framework to (a) different transition weight strategies, (b) biases in the pretrained classifier, and (c) the number of parallel samples acquired during training.}
\label{fig: ablation}
\end{figure*}

\section{Ablation Studies}

To validate our model design choices and assess the sensitivity of our framework to key hyperparameters, we conduct comprehensive ablation studies on ACL sprain diagnosis. We analyze three critical components: (a) Transition Weight (b) Pretrained Classifier Bias and (c) Parallel Samples, summarized in Figure~\ref{fig: ablation}, provide insights into the robustness and design rationale of our approach.

\subsection{Sensitivity to Step-wise Transition Weight}

We investigate five different transition strategies for balancing disease detection and severity quantification rewards during the sampling process. As shown in Figure~\ref{fig: ablation}(a), the policy without any transition mechanism, which we assign equal weight to both objectives, performs poorly, demonstrating the necessity of structured reward scheduling with lexicographic ordering. While both linear interpolation and standard cosine interpolation show improvement over the no-transition baseline, our proposed cosine interpolation achieves superior performance compared to linear interpolation by enabling adaptive and customized transition as shown in Figure~\ref{fig: adaptive weight} instead of fixed-step transition. 
For the ACL Sprain diagnosis task, this configuration achieved the highest AUC with $\beta = 0.2$ (early transition at step $34$) and demonstrated efficient convergence compared to $\beta = 0$ (transition at step $40$) or $\beta = -0.2$ (late transition at step $45$). 
This result may be attributable to the lower relative complexity of detecting disease compared to quantifying its severity. This confirms that a carefully designed transition enhances diagnostic performance and suggests that task-specific transition points could be beneficial, potentially learned as a function of task complexity.

\subsection{Sensitivity to Pretrained Classifier Performance}
\label{sec: pretrained_classifier}
Our framework employs pretrained classifiers as both reward models and inference operators, which may lead to error propagation to the sampling policy. To assess this potential, we conduct sensitivity analysis with two simulated bias conditions: (i) \textbf{Class-imbalance bias}: We train classifiers using standard cross-entropy loss instead of the weighted cross-entropy employed in our main approach, simulating scenarios where class imbalance is not properly addressed. (ii) \textbf{Label annotation bias}: We randomly flip 10\% of severity labels in the training set to simulate human annotation errors commonly encountered in clinical datasets.

As illustrated in Figure~\ref{fig: ablation}(b), As illustrated in Figure~\ref{fig: ablation}(b), both bias conditions affect policy performance. Label annotation bias results in 7\% AUC degradation, as the policy receives noisy reward signals from erroneous severity predictions. Class-imbalance bias leads to a 35\% performance decrease, indicating that the policy becomes trapped by the majority class distribution due to the biased reward. Our method still does better than using random sampling.

These findings underscore the critical importance of reward quality in reinforcement learning frameworks, which is widely known in the RL community. They also demark the criticality of class-imbalance as a worse form of bias, which is expected as this is a consistent bias towards the population vs a random effect (such as label bias). We highlight that the classifier used in our experiments was using bias mitigation strategies precisely for these reasons.

\subsection{Sensitivity to Parallel Acquisition Rate}

We examine the effect of parallel acquisition number (\(q\)) on policy training efficiency and final performance. As shown in Figure~\ref{fig: ablation}(c), we compare our chosen setting (\(q=8\)) with lower (\(q=2\)) and higher (\(q=16\)) parallel acquisition rates. With fewer parallel samples (\(q=2\)), the policy exhibits slower convergence and plateaus at a later sampling step, matching the performance of \(q=8\) only after extensive sampling. This indicates that insufficient parallel acquisition constrains action space exploration during training, hindering the agent's ability to efficiently identify diagnostically relevant $k$-space features. Conversely, with more parallel samples (\(q=16\)), the policy achieves slightly earlier plateauing compared to \(q=8\) but does not surpass the final performance level. This suggests that while higher parallel acquisition rates can accelerate the discovery of diagnostic features, they provide diminishing returns beyond an optimal threshold. Our chosen value of \(q=8\) represents an effective balance between exploration efficiency and computational resources.

\section{Discussion}     

\subsection{Clinical Relevance}

Our framework enables the efficient acquisition of diagnostically relevant $k$-space data while sequentially performing disease detection and severity quantification in a single scanning session, such as stroke assessment (detecting a bleed first, then quantifying its volume or location)~\citep{2022Guideline} or acute trauma diagnosis, where a logical, sequential workflow is standard practice~\citep{kool2007advanced}. This approach enables comprehensive assessment during the scanning process, potentially eliminating the need for follow-up examinations. The sequential diagnosis paradigm completes multi-objective diagnostic tasks within a single scanning session without requiring full $k$-space acquisition, thereby significantly reducing acquisition time. 

The acquisition acceleration of the proposed pipeline via patient-level active sampling could substantially enhance the clinical utility of low-field MRI, making them more viable for PoC applications~\citep{geethanath2019accessible}. We envision practical implementations in preliminary screening scenarios where rapid assessments are crucial to guide resource allocation decisions~\citep{brady2021radiology}, which is particularly valuable in settings with limited healthcare infrastructure or in emergency medicine contexts~\citep{mazurek2021portable} where time-critical diagnostic information is essential.

\subsection{Limitations and Future Works} 

Our method has been implemented as a proof-of-concept on the fastMRI knee dataset, focusing on specific pathologies where $k$-space data are publicly available with annotated disease and severity labels, proving the feasibility of our method in in-site, rapid assessment. Future work could explore other applications such as stroke~\citep{ho2019predicting} or acute brain trauma diagnosis~\citep{mazurek2021portable}, where rapid assessment from limited data is crucial. Although such datasets are not publicly available, we hope our open-source code will encourage researchers with access to evaluate our method in these critical settings.

Methodologically, our approach is tailored to problems with a known and fixed diagnostic sequence, and therefore may not be suitable for clinical scenarios where multiple diagnostic tasks are truly parallel or where the optimal order of tasks is unknown. Future iterations could explore dynamic policies that learn not only \textit{what} to sample but also the optimal \textit{order} of diagnostic objectives. Additionally, while our current hard-gating mechanism is effective for this proof-of-concept, clinical deployment would benefit from more granular uncertainty quantification. Future research could explore uncertainty-aware decision making, where methods like Monte Carlo dropout dynamically modulate the diagnosis threshold, effectively adding a safety layer for ambiguous cases.

A critical area for future improvement is robustness to model bias. As analyzed in Section~\ref{sec: pretrained_classifier}, our reliance on pre-trained classifiers can propagate inherent biases and leave the policy vulnerable to domain shift~\citep{wachinger2021detect}. Nevertheless, our policy-guided sampling can mitigate the bias-induced performance drop compared to random sampling as shown in Figure~\ref{fig: ablation}(b), motivating future work on bias-aware policy training. Promising directions include integrating constraints directly into the reward function or employing adversarial debiasing techniques to suppress shortcut attributes (e.g., acquisition site or protocol).

Finally, we acknowledge the hardware constraints associated with practical deployment. The real-time feedback loop required for dynamic, line-by-line trajectory control is not a standard feature on most current mainstream scanners. However, previous works~\citep{pub.1134187574, liu2024rl} demonstrate the feasibility of autonomous, closed-loop $k$-space trajectory on a clinical MRI scanner with real-time feedback to dynamically update the acquisition sequence. Our research builds on these pioneering efforts, serving as a proof-of-concept to motivate the broader development of programmable acquisition systems that can support intelligent sampling.

\section{Conclusion}
We propose a novel MORL framework for sequential diagnostic assessment directly from undersampled $k$-space data. By structuring clinical diagnosis as a sequential decision process with lexicographic ordering, our method achieves diagnostic performance comparable to fully sampled references and outperforms undersampled baselines while significantly reducing acquisition time.
Empirical results on two tasks show that our diagnosis-oriented sampling strategy surpasses reconstruction-based methods in both diagnostic accuracy and efficiency. Compared to single-objective diagnostic samplers, our approach achieves substantial $k$-space savings within a single scan without compromising performance.
Our step-wise weighting mechanism effectively prioritizes disease detection early in the acquisition before transitioning to severity assessment, thereby aligning with clinical workflows. Furthermore, the convergence of sampling trajectories across tasks supports the existence of shared informative $k$-space regions, enabling efficient, comprehensive assessment in a single session. This work advances direct $k$-space inference, paving the way for clinically viable, PoC MRI diagnostics with reduced acquisition demands.

\acks{This work was supported in part by National Institutes of Health (NIH) grant 7R01HL148788-03. Yuning Du thanks additional financial support from the School of Engineering, the University of Edinburgh. 
Sotirios A.\ Tsaftaris acknowledges support from the Royal Academy of Engineering and the Research Chairs and Senior Research Fellowships scheme (grant RCSRF1819\textbackslash8\textbackslash 25), and the UK Engineering and Physical Sciences Research Council (EPSRC) support via grant EP/X017680/1, and the UKRI AI programme, Causality in Healthcare AI Hub (CHAI, grant EP/Y028856/1).}

%
\ethics{The work follows appropriate ethical standards in conducting research and writing the manuscript, following all applicable laws and regulations regarding treatment of animals or human subjects.}

\coi{The authors declare that they have no known competing financial interests or personal relationships that could have appeared to influence the work reported in this paper.}

\data{Authors submitting articles to \textsc{Melba} are required to include a Data Availability Statement in their manuscripts. The Data Availability Statement should clearly indicate whether the data supporting the findings of the study are available and, if so, how readers can access them. If the data are not available, authors should provide a brief justification for not sharing the data.}

\bibliography{reference.bib}

@article{lin2021artificial,
  title={Artificial intelligence for MR image reconstruction: an overview for clinicians},
  author={Lin, Dana J and Johnson, Patricia M and Knoll, Florian and Lui, Yvonne W},
  journal={Journal of Magnetic Resonance Imaging},
  volume={53},
  number={4},
  pages={1015--1028},
  year={2021},
  publisher={Wiley Online Library}
}

@article{wu2024learning,
  title={Learning task-specific strategies for accelerated MRI},
  author={Wu, Zihui and Yin, Tianwei and Sun, Yu and Frost, Robert and van der Kouwe, Andre and Dalca, Adrian V and Bouman, Katherine L},
  journal={IEEE transactions on computational imaging},
  volume={10},
  pages={1040--1054},
  year={2024},
  publisher={IEEE}
}

@article{aggarwal2020j,
  title={J-MoDL: Joint model-based deep learning for optimized sampling and reconstruction},
  author={Aggarwal, Hemant Kumar and Jacob, Mathews},
  journal={IEEE journal of selected topics in signal processing},
  volume={14},
  number={6},
  pages={1151--1162},
  year={2020},
  publisher={IEEE}
}

@inproceedings{alkan2020learning,
  title={Learning to sample MRI via variational information maximization},
  author={Alkan, Cagan and Mardani, Morteza and Vasanawala, Shreyas and Pauly, John M},
  booktitle={NeurIPS 2020 Workshop on Deep Learning and Inverse Problems},
  year={2020}
}

@article{arastu2015prevalence,
  title={Prevalence and consequences of delayed diagnosis of anterior cruciate ligament ruptures},
  author={Arastu, MH and Grange, S and Twyman, R},
  journal={Knee Surgery, Sports Traumatology, Arthroscopy},
  volume={23},
  number={4},
  pages={1201--1205},
  year={2015},
  publisher={Springer}
}

@inproceedings{chaithya2022hybrid,
  title={Hybrid learning of non-cartesian k-space trajectory and MR image reconstruction networks},
  author={Chaithya, GR and Ramzi, Zaccharie and Ciuciu, Philippe},
  booktitle={2022 IEEE 19th International Symposium on Biomedical Imaging (ISBI)},
  pages={1--5},
  year={2022},
  organization={IEEE}
}

@article{kool2007advanced,
  title={Advanced Trauma Life Support{\textregistered}. ABCDE from a radiological point of view},
  author={Kool, Digna R and Blickman, Johan G},
  journal={Emergency radiology},
  volume={14},
  number={3},
  pages={135--141},
  year={2007},
  publisher={Springer}
}

@article{2022Guideline,
title = "2022 Guideline for the Management of Patients With Spontaneous Intracerebral Hemorrhage: A Guideline From the American Heart Association/American Stroke Association",
author = "Greenberg, {Steven M.} and Ziai, {Wendy C.} and Charlotte Cordonnier and Dar Dowlatshahi and Brandon Francis and Goldstein, {Joshua N.} and Hemphill, {J. Claude} and Ronda Johnson and Keigher, {Kiffon M.} and Mack, {William J.} and J. Mocco and Newton, {Eileena J.} and Ruff, {Ilana M.} and Sansing, {Lauren H.} and Sam Schulman and Selim, {Magdy H.} and Sheth, {Kevin N.} and Nikola Sprigg and Sunnerhagen, {Katharina S.}",
year = "2022",
month = jul,
day = "1",
doi = "10.1161/STR.0000000000000407",
language = "English (US)",
volume = "53",
pages = "E282--E361",
journal = "Stroke",
issn = "0039-2499",
publisher = "Wolters Kluwer Health",
number = "7",
}

@article{martinini2022deep,
  title={A deep learning method for optimal undersampling patterns and image recovery for MRI exploiting losses and projections},
  author={Martinini, Filippo and Mangia, Mauro and Marchioni, Alex and Rovatti, Riccardo and Setti, Gianluca},
  journal={IEEE Journal of Selected Topics in Signal Processing},
  volume={16},
  number={4},
  pages={713--724},
  year={2022},
  publisher={IEEE}
}

@inproceedings{li2024classification,
  title={Classification, Regression and Segmentation directly from k-Space in Cardiac MRI},
  author={Li, Ruochen and Pan, Jiazhen and Zhu, Youxiang and Ni, Juncheng and Rueckert, Daniel},
  booktitle={International Workshop on Machine Learning in Medical Imaging},
  pages={31--41},
  year={2024},
  organization={Springer}
}

@article{reina2025mrilow,
title = "{MRI} at low-field: a review of software solutions for improving {SNR}",
author = "Reina Ayde and Marc Vornehm and Yujiao Zhao and Florian Knoll and Wu, {Ed X.} and Mathieu Sarracanie",
year = "2025",
month = jan,
doi = "10.1002/nbm.5268",
language = "English",
volume = "38",
journal = "NMR in Biomedicine",
issn = "0952-3480",
publisher = "John Wiley and Sons Ltd",
number = "1",
}

@article{pub.1134187574,
 author = {Contijoch, Francisco and Han, Yuchi and Iyer, Srikant Kamesh and Kellman, Peter and Gualtieri, Gene and Elliott, Mark A. and Berisha, Sebastian and Gorman, Joseph H. and Gorman, Robert C. and Pilla, James J. and Witschey, Walter R. T.},
 doi = {10.1371/journal.pone.0244286},
 journal = {PLOS ONE},
 number = {12},
 pages = {e0244286},
 title = {Closed-loop control of k-space sampling via physiologic feedback for cine MRI},
 volume = {15},
 year = {2020}
}

@article{liu2024rl,
       author = {{Liu}, Yiming and {Pang}, Yanwei and {Jin}, Ruiqi and {Hou}, Yonghong and {Li}, Xuelong},
        title = "{Reinforcement Learning and Transformer for Fast Magnetic Resonance Imaging Scan}",
      journal = {IEEE Transactions on Emerging Topics in Computational Intelligence},
         year = 2024,
        month = jan,
       volume = {8},
       number = {3},
        pages = {2310-2323},
          doi = {10.1109/TETCI.2024.3358180},
       adsurl = {https://ui.adsabs.harvard.edu/abs/2024ITECI...8.2310L}
}

@article{wang2023efficient,
  title={Efficient approximation of Jacobian matrices involving a non-uniform fast Fourier transform (NUFFT)},
  author={Wang, Guanhua and Fessler, Jeffrey A},
  journal={IEEE transactions on computational imaging},
  volume={9},
  pages={43--54},
  year={2023},
  publisher={IEEE}
}

@article{kingma2014adam,
  title={Adam: A method for stochastic optimization},
  author={Kingma, Diederik P and Ba, Jimmy},
  journal={arXiv preprint arXiv:1412.6980},
  year={2014}
}

@InProceedings{pmlr-v202-mao23b,
  title = 	 {Cross-Entropy Loss Functions: Theoretical Analysis and Applications},
  author =       {Mao, Anqi and Mohri, Mehryar and Zhong, Yutao},
  booktitle = 	 {Proceedings of the 40th International Conference on Machine Learning},
  pages = 	 {23803--23828},
  year = 	 {2023},
  volume = 	 {202},
  series = 	 {Proceedings of Machine Learning Research},
  publisher =    {PMLR},
}

@article{ravula2023optimizing,
  title={Optimizing sampling patterns for compressed sensing MRI with diffusion generative models},
  author={Ravula, Sriram and Levac, Brett and Jalal, Ajil and Tamir, Jonathan I and Dimakis, Alexandros G},
  journal={arXiv preprint arXiv:2306.03284},
  year={2023}
}

@article{cosineinterpolation,
  author={Maeland, E.},
  journal={IEEE Transactions on Medical Imaging}, 
  title={On the comparison of interpolation methods}, 
  year={1988},
  volume={7},
  number={3},
  pages={213-217},
  doi={10.1109/42.7784}}

@article{geethanath2019accessible,
  title={Accessible magnetic resonance imaging: a review},
  author={Geethanath, Sairam and Vaughan Jr, John Thomas},
  journal={Journal of Magnetic Resonance Imaging},
  volume={49},
  number={7},
  pages={e65--e77},
  year={2019},
  publisher={Wiley Online Library}
}

@article{zbontar2018fastmri,
  title={fastMRI: An open dataset and benchmarks for accelerated MRI},
  author={Zbontar, Jure and Knoll, Florian and Sriram, Anuroop and Murrell, Tullie and Huang, Zhengnan and Muckley, Matthew J and Defazio, Aaron and Stern, Ruben and Johnson, Patricia and Bruno, Mary and others},
  journal={arXiv preprint arXiv:1811.08839},
  year={2018}
}

@book{sondik1971optimal,
  title={The optimal control of partially observable Markov processes},
  author={Sondik, Edward Jay},
  year={1971},
  publisher={Stanford University}
}

@article{liu2014multiobjective,
  title={Multiobjective reinforcement learning: A comprehensive overview},
  author={Liu, Chunming and Xu, Xin and Hu, Dewen},
  journal={IEEE Transactions on Systems, Man, and Cybernetics: Systems},
  volume={45},
  number={3},
  pages={385--398},
  year={2014},
  publisher={IEEE}
}

@article{skalse2022lexicographic,
  title={Lexicographic multi-objective reinforcement learning},
  author={Skalse, Joar and Hammond, Lewis and Griffin, Charlie and Abate, Alessandro},
  journal={arXiv preprint arXiv:2212.13769},
  year={2022}
}

@inproceedings{ronneberger2015u,
  title={U-net: Convolutional networks for biomedical image segmentation},
  author={Ronneberger, Olaf and Fischer, Philipp and Brox, Thomas},
  booktitle={Medical image computing and computer-assisted intervention--MICCAI 2015: 18th international conference, Munich, Germany, October 5-9, 2015, proceedings, part III 18},
  pages={234--241},
  year={2015},
  organization={Springer}
}

@inproceedings{he2016deep,
  title={Deep residual learning for image recognition},
  author={He, Kaiming and Zhang, Xiangyu and Ren, Shaoqing and Sun, Jian},
  booktitle={Proceedings of the IEEE conference on computer vision and pattern recognition},
  pages={770--778},
  year={2016}
}

@article{zhao2022fastmri+,
  title={fastMRI+, clinical pathology annotations for knee and brain fully sampled magnetic resonance imaging data},
  author={Zhao, Ruiyang and Yaman, Burhaneddin and Zhang, Yuxin and Stewart, Russell and Dixon, Austin and Knoll, Florian and Huang, Zhengnan and Lui, Yvonne W and Hansen, Michael S and Lungren, Matthew P},
  journal={Scientific Data},
  volume={9},
  number={1},
  pages={152},
  year={2022},
  publisher={Nature Publishing Group UK London}
}

@article{bahadir2020deep,
  title={Deep-learning-based optimization of the under-sampling pattern in MRI},
  author={Bahadir, Cagla D and Wang, Alan Q and Dalca, Adrian V and Sabuncu, Mert R},
  journal={IEEE Transactions on Computational Imaging},
  volume={6},
  pages={1139--1152},
  year={2020},
  publisher={IEEE}
}

@article{wang2004image,
  title={Image quality assessment: from error visibility to structural similarity},
  author={Wang, Zhou and Bovik, Alan C and Sheikh, Hamid R and Simoncelli, Eero P},
  journal={IEEE transactions on image processing},
  volume={13},
  number={4},
  pages={600--612},
  year={2004},
  publisher={IEEE}
}

@misc{kool2019buy,
title={Buy 4 {REINFORCE} Samples, Get a Baseline for Free!},
author={Wouter Kool and Herke van Hoof and Max Welling},
year={2019},
url={https://openreview.net/forum?id=r1lgTGL5DE}
}

@inproceedings{vlontzos2019multiple,
  title={Multiple landmark detection using multi-agent reinforcement learning},
  author={Vlontzos, Athanasios and Alansary, Amir and Kamnitsas, Konstantinos and Rueckert, Daniel and Kainz, Bernhard},
  booktitle={Medical Image Computing and Computer Assisted Intervention--MICCAI 2019: 22nd International Conference, Shenzhen, China, October 13--17, 2019, Proceedings, Part IV 22},
  pages={262--270},
  year={2019},
  organization={Springer}
}

@article{zhang2021sequential,
  title={Sequential conditional reinforcement learning for simultaneous vertebral body detection and segmentation with modeling the spine anatomy},
  author={Zhang, Dong and Chen, Bo and Li, Shuo},
  journal={Medical image analysis},
  volume={67},
  pages={101861},
  year={2021},
  publisher={Elsevier}
}

@article{hori2021low,
  title={Low-field magnetic resonance imaging: its history and renaissance},
  author={Hori, Masaaki and Hagiwara, Akifumi and Goto, Masami and Wada, Akihiko and Aoki, Shigeki},
  journal={Investigative radiology},
  volume={56},
  number={11},
  pages={669--679},
  year={2021},
  publisher={LWW}
}

@techreport{royal2018standards,
  title = {Standards for interpretation and reporting of imaging investigations},
  author = {{Royal College of Radiologists}},
  institution = {The Royal College of Radiologists},
  address = {London, UK},
  year = {2018},
  edition = {Second},
}

@misc{anzai2022point,
  title={Point-of-Care Low-Field-Strength MRI Is Moving Beyond the Hype},
  author={Anzai, Yoshimi and Moy, Linda},
  journal={Radiology},
  volume={305},
  number={3},
  pages={672--673},
  year={2022},
  publisher={Radiological Society of North America}
}

@article{brady2021radiology,
  title={Radiology in the era of value-based healthcare: a multi-society expert statement from the ACR, CAR, ESR, IS3R, RANZCR, and RSNA},
  author={Brady, Adrian P and Bello, Jaqueline A and Derchi, Lorenzo E and Fuchsj{\"a}ger, Michael and Goergen, Stacy and Krestin, Gabriel P and Lee, Emil JY and Levin, David C and Pressacco, Josephine and Rao, Vijay M and others},
  journal={Canadian Association of Radiologists Journal},
  volume={72},
  number={2},
  pages={208--214},
  year={2021},
  publisher={SAGE Publications Sage CA: Los Angeles, CA}
}

@article{wachinger2021detect,
  title={Detect and correct bias in multi-site neuroimaging datasets},
  author={Wachinger, Christian and Rieckmann, Anna and P{\"o}lsterl, Sebastian and Alzheimer’s Disease Neuroimaging Initiative and others},
  journal={Medical Image Analysis},
  volume={67},
  pages={101879},
  year={2021},
  publisher={Elsevier}
}

@article{yu2021reinforcement,
  title={Reinforcement learning in healthcare: A survey},
  author={Yu, Chao and Liu, Jiming and Nemati, Shamim and Yin, Guosheng},
  journal={ACM Computing Surveys (CSUR)},
  volume={55},
  number={1},
  pages={1--36},
  year={2021},
  publisher={ACM New York, NY}
}

@inproceedings{liao2020iteratively,
  title={Iteratively-refined interactive 3D medical image segmentation with multi-agent reinforcement learning},
  author={Liao, Xuan and Li, Wenhao and Xu, Qisen and Wang, Xiangfeng and Jin, Bo and Zhang, Xiaoyun and Wang, Yanfeng and Zhang, Ya},
  booktitle={Proceedings of the IEEE/CVF conference on computer vision and pattern recognition},
  pages={9394--9402},
  year={2020}
}

@article{huang2021deep,
  title={A deep multi-task learning framework for brain tumor segmentation},
  author={Huang, He and Yang, Guang and Zhang, Wenbo and Xu, Xiaomei and Yang, Weiji and Jiang, Weiwei and Lai, Xiaobo},
  journal={Frontiers in oncology},
  volume={11},
  pages={690244},
  year={2021},
  publisher={Frontiers Media SA}
}

@article{ho2019predicting,
  title={Predicting ischemic stroke tissue fate using a deep convolutional neural network on source magnetic resonance perfusion images},
  author={Ho, King Chung and Scalzo, Fabien and Sarma, Karthik V and Speier, William and El-Saden, Suzie and Arnold, Corey},
  journal={Journal of Medical Imaging},
  volume={6},
  number={2},
  pages={026001--026001},
  year={2019},
  publisher={Society of Photo-Optical Instrumentation Engineers}
}

@article{mazurek2021portable,
  title={Portable, bedside, low-field magnetic resonance imaging for evaluation of intracerebral hemorrhage},
  author={Mazurek, Mercy H and Cahn, Bradley A and Yuen, Matthew M and Prabhat, Anjali M and Chavva, Isha R and Shah, Jill T and Crawford, Anna L and Welch, E Brian and Rothberg, Jonathan and Sacolick, Laura and others},
  journal={Nature communications},
  volume={12},
  number={1},
  pages={5119},
  year={2021},
  publisher={Nature Publishing Group UK London}
}

@inproceedings{xuan2020learning,
  title={Learning MRI k-space subsampling pattern using progressive weight pruning},
  author={Xuan, Kai and Sun, Shanhui and Xue, Zhong and Wang, Qian and Liao, Shu},
  booktitle={Medical Image Computing and Computer Assisted Intervention--MICCAI 2020: 23rd International Conference, Lima, Peru, October 4--8, 2020, Proceedings, Part II 23},
  pages={178--187},
  year={2020},
  organization={Springer}
}

@article{wang2022joint,
  title={Joint optimization of Cartesian sampling patterns and reconstruction for single-contrast and multi-contrast fast magnetic resonance imaging},
  author={Wang, Jiechao and Yang, Qinqin and Yang, Qizhi and Xu, Lina and Cai, Congbo and Cai, Shuhui},
  journal={Computer Methods and Programs in Biomedicine},
  volume={226},
  pages={107150},
  year={2022},
  publisher={Elsevier}
}

@inproceedings{zhang2020extending,
  title={Extending LOUPE for K-space Under-sampling Pattern Optimization in Multi-coil MRI},
  author={Zhang, Jinwei and Zhang, Hang and Wang, Alan and Zhang, Qihao and Sabuncu, Mert and Spincemaille, Pascal and Nguyen, Thanh D and Wang, Yi},
  booktitle={Machine Learning for Medical Image Reconstruction: Third International Workshop, MLMIR 2020, Held in Conjunction with MICCAI 2020, Lima, Peru, October 8, 2020, Proceedings 3},
  pages={91--101},
  year={2020},
  organization={Springer}
}

@inproceedings{han2024advancing,
  title={Advancing text-driven chest x-ray generation with policy-based reinforcement learning},
  author={Han, Woojung and Kim, Chanyoung and Ju, Dayun and Shim, Yumin and Hwang, Seong Jae},
  booktitle={International Conference on Medical Image Computing and Computer-Assisted Intervention},
  pages={56--66},
  year={2024},
  organization={Springer}
}

@article{bakker2020experimental,
  title="{Experimental design for MRI by greedy policy search}",
  author={Bakker, Tim and van Hoof, Herke and Welling, Max},
  journal={Advances in Neural Information Processing Systems},
  volume={33},
  pages={18954--18966},
  year={2020}
}

@article{yen2024adaptive,
  title={Adaptive sampling of k-space in magnetic resonance for rapid pathology prediction},
  author={Yen, Chen-Yu and Singhal, Raghav and Sharma, Umang and Ranganath, Rajesh and Chopra, Sumit and Pinto, Lerrel},
  journal={arXiv preprint arXiv:2406.04318},
  year={2024}
}

@inproceedings{zhang2019reducing,
  title={Reducing uncertainty in undersampled MRI reconstruction with active acquisition},
  author={Zhang, Zizhao and Romero, Adriana and Muckley, Matthew J and Vincent, Pascal and Yang, Lin and Drozdzal, Michal},
  booktitle={Proceedings of the IEEE/CVF conference on computer vision and pattern recognition},
  pages={2049--2058},
  year={2019}
}

@inproceedings{pineda2020active,
  title={Active MR k-space sampling with reinforcement learning},
  author={Pineda, Luis and Basu, Sumana and Romero, Adriana and Calandra, Roberto and Drozdzal, Michal},
  booktitle={International Conference on Medical Image Computing and Computer-Assisted Intervention},
  pages={23--33},
  year={2020},
  organization={Springer}
}

@inproceedings{du2024mri,
  title="{The MRI scanner as a diagnostic: image-less active sampling}",
  author={Du, Yuning and Dharmakumar, Rohan and Tsaftaris, Sotirios A},
  booktitle={International Conference on Medical Image Computing and Computer-Assisted Intervention},
  pages={467--476},
  year={2024},
  organization={Springer}
}

@article{singhal2023feasibility,
  title={On the feasibility of machine learning augmented magnetic resonance for point-of-care identification of disease},
  author={Singhal, Raghav and Sudarshan, Mukund and Mahishi, Anish and Kaushik, Sri and Ginocchio, Luke and Tong, Angela and Chandarana, Hersh and Sodickson, Daniel K and Ranganath, Rajesh and Chopra, Sumit},
  journal={arXiv preprint arXiv:2301.11962},
  year={2023}
}

@article{zhao2023multi,
  title={Multi-task deep learning for medical image computing and analysis: A review},
  author={Zhao, Yan and Wang, Xiuying and Che, Tongtong and Bao, Guoqing and Li, Shuyu},
  journal={Computers in Biology and Medicine},
  volume={153},
  pages={106496},
  year={2023},
  publisher={Elsevier}
}

@article{wang2022b,
  title={B-spline parameterized joint optimization of reconstruction and k-space trajectories (bjork) for accelerated 2d mri},
  author={Wang, Guanhua and Luo, Tianrui and Nielsen, Jon-Fredrik and Noll, Douglas C and Fessler, Jeffrey A},
  journal={IEEE Transactions on Medical Imaging},
  volume={41},
  number={9},
  pages={2318--2330},
  year={2022},
  publisher={IEEE}
}

@inproceedings{schlemper2018cardiac,
  title={Cardiac MR segmentation from undersampled k-space using deep latent representation learning},
  author={Schlemper, Jo and Oktay, Ozan and Bai, Wenjia and Castro, Daniel C and Duan, Jinming and Qin, Chen and Hajnal, Jo V and Rueckert, Daniel},
  booktitle={Medical Image Computing and Computer Assisted Intervention--MICCAI 2018: 21st International Conference, Granada, Spain, September 16-20, 2018, Proceedings, Part I},
  pages={259--267},
  year={2018},
  organization={Springer}
}

@article{yiasemis2024retrospective,
  title={On retrospective k-space subsampling schemes for deep MRI reconstruction},
  author={Yiasemis, George and S{\'a}nchez, Clara I and Sonke, Jan-Jakob and Teuwen, Jonas},
  journal={Magnetic Resonance Imaging},
  volume={107},
  pages={33--46},
  year={2024},
  publisher={Elsevier}
}

@article{zeng2021review,
  title={A review on deep learning MRI reconstruction without fully sampled k-space},
  author={Zeng, Gushan and Guo, Yi and Zhan, Jiaying and Wang, Zi and Lai, Zongying and Du, Xiaofeng and Qu, Xiaobo and Guo, Di},
  journal={BMC Medical Imaging},
  volume={21},
  number={1},
  pages={195},
  year={2021},
  publisher={Springer}
}





\end{document}